\documentclass[lettersize,journal]{IEEEtran}
\usepackage{amsmath,amsfonts}
\usepackage{algorithmic}
\usepackage{algorithm}
\usepackage{array}
\usepackage[caption=false,font=normalsize,labelfont=sf,textfont=sf]{subfig}
\usepackage{textcomp}
\usepackage{stfloats}
\usepackage{url}
\usepackage{verbatim}
\usepackage{graphicx}
\usepackage{cite}
\usepackage{amssymb}
\begin{document}
	\bstctlcite{IEEEexample:BSTcontrol}
	\title{Fluid Antenna-Aided Robust Secure Uplink Transmission in LEO Satellite Constellations}
	
	\author{Suhong Luo, Pan Tang, Jianhua Zhang, \IEEEmembership{Fellow,~IEEE}, Ji Wang, \IEEEmembership{Senior Member,~IEEE}, Yixuan Li, \\ Xingwang Li, \IEEEmembership{Senior Member,~IEEE}
		\thanks{\textit{(Corresponding author: Pan Tang.)}
			
		Suhong Luo, Pan Tang, and Jianhua Zhang are with the State Key Laboratory of Networking and Switching Technology, Beijing University of Posts and Telecommunications, Beijing 100876, China (e-mail: shl@bupt.edu.cn; tangpan27@bupt.edu.cn; jhzhang@bupt.edu.cn).
		
		Ji Wang and Yixuan Li are with the Department of Electronics and Information Engineering, Central China Normal University, Wuhan 430079, China (e-mail: jiwang@ccnu.edu.cn; yixuanli@mails.ccnu.edu.cn).		
		
		Xingwang Li is with the School of Physics and Electronic Information Engineering, Henan Polytechnic University, Jiaozuo 454003, China (e-mail: lixingwang@hpu.edu.cn).
	}}
	\markboth{Journal of \LaTeX\ Class Files,~Vol.~XX, No.~XX, November~2026}%
	{Shell \MakeLowercase{\textit{et al.}}: A Sample Article Using IEEEtran.cls for IEEE Journals}
	
	
	\maketitle
	
	\begin{abstract}
		Secure uplink transmission in low-earth orbit (LEO) satellite constellations faces significant challenges due to the time-varying spatial distributions of serving and potential eavesdropping satellites, particularly when their angular separations are small and eavesdropper channel state information (CSI) is imperfect. To address these challenges, this paper investigates fluid antenna (FA)-aided robust secure uplink transmission in LEO satellite constellations. To avoid frequent antenna reconfiguration, the FA positions are optimized over an observation interval and remain fixed throughout the interval, while transmit beamforming dynamically adapts to the time-varying satellite channels. We formulate an average worst-case secrecy rate maximization problem by jointly optimizing FA positions and time-varying beamforming under norm-bounded eavesdropper CSI uncertainties and per-slot secrecy rate constraints. Two new alternating optimization (AO) algorithms are developed to solve the resulting non-convex problem. For robust beamforming optimization, the S-procedure, semidefinite relaxation (SDR), and first-order approximation are employed to construct a tractable convex problem, followed by Gaussian randomization for rank-one recovery. For FA position optimization, we derive concave surrogate functions for successive convex approximation (SCA) and develop a differential evolution (DE)-based search strategy to reduce computational complexity. Numerical results demonstrate that both FA-aided schemes consistently outperform conventional fixed-position antenna arrays across different constellation configurations and CSI uncertainty levels. Moreover, the two algorithms exhibit different performance-complexity tradeoffs as the number of FA elements increases, with DE offering a computational advantage for larger arrays while SCA can achieve higher secrecy rates.
	\end{abstract}
	
	\begin{IEEEkeywords}
		6G, LEO satellite constellations, fluid antenna system, physical layer security, transmit beamforming, antenna position optimization.
	\end{IEEEkeywords}

	\section{Introduction}
	
	Low-earth orbit (LEO) satellite communications offer shorter propagation delays and lower free-space path loss than satellite systems in higher orbits, making them a promising component of sixth-generation (6G) non-terrestrial networks (NTNs) \cite{kodheli2020satellite,di2019ultra}. Owing to orbital motion, an individual LEO satellite remains visible to a given ground station (GS) for only a limited period. Constellations distributed across multiple orbital planes enable successive satellite access, supporting sustained ground-to-satellite connectivity. As of October 15, 2026, Starlink and OneWeb had 11154 and 654 satellites in orbit, respectively \cite{mcdowell_starlink_stats}. The growing scale of LEO constellations also raises concerns about uplink security. When a GS transmits confidential information to its serving satellite, multiple other satellites may be simultaneously visible. For example, an ephemeris-based study reported that more than 50 Starlink satellites could be geometrically visible in certain non-polar regions at a minimum elevation angle of $10^\circ$ \cite{kozhaya2024opportunistic}. Although visibility alone does not imply interception capability, satellites equipped with compatible receivers may act as potential eavesdroppers. In particular, an eavesdropping satellite located close to the serving satellite in angular separation may receive substantial gain from the GS transmit beam, making it difficult to suppress information leakage through directional transmission alone. Moreover, the continuous orbital motion of satellites changes their relative angular positions, introducing time-varying eavesdropping threats to secure uplink communications.
		
	\subsection{Related Works}
	
	Physical-layer security protects confidential information by exploiting differences between legitimate and eavesdropping channels \cite{mukherjee2014principles}. Existing approaches include secure beamforming and power allocation \cite{li2011optimal,liao2011qos,zhou2010secure}, artificial noise and cooperative jamming \cite{goel2008guaranteeing,dong2010improving,mukherjee2011robust}, and reconfigurable intelligent surfaces (RISs) \cite{cui2019secure,shen2019secrecy,guan2020intelligent,yu2020robust}. However, these approaches generally assume fixed-position transmit antenna arrays, in which the element locations and inter-element spacings remain unchanged. Beamforming weights can be optimized to balance the desired signal gain against information leakage, but the array geometry cannot be adjusted to improve spatial discrimination. In the LEO satellite uplink communication systems, the serving satellite and potential eavesdroppers may be closely spaced in angle as viewed from the GS, resulting in highly correlated steering vectors. Suppressing the signal toward an eavesdropper may therefore also reduce the gain toward the serving satellite.Fluid antenna (FA) technology provides additional spatial degrees of freedom through flexible port selection or antenna position adjustment within a prescribed region \cite{wu2025fluid}. By adjusting the antenna positions, FA arrays can modify their spatial responses, allowing joint optimization of array geometry and beamforming weights to enhance directional selectivity.
	
	FA technology has attracted considerable attention for improving wireless communication performance through antenna position optimization. The authors in \cite{zhou2025fluid} jointly optimizes transmit beamforming and FA positions at the base station and receivers to maximize the information rate while satisfying an energy-harvesting requirement in a simultaneous wireless information and power transfer systems. In \cite{cheng2024FAsumrate}, the transmit FA positions and beamforming vectors were jointly optimized to maximize the multiuser sum rate. For near-field communications, \cite{chen2025FAnearfield} investigated energy efficiency maximization by jointly optimizing transmit beamforming and receive FA positions through an alternating optimization (AO) algorithm. The authors in \cite{tang2025FAuplink} considered an uplink system equipped with FAs at both the transmitter and receiver sides, where transmit power and antenna positions were optimized to maximize system capacity. More recently, \cite{li2026FApathfollowing} developed a path-following optimization method for joint FA placement and beamforming design, addressing max-min rate and sum-capacity optimization in multiuser downlink systems. These studies have established effective approaches for exploiting FA position reconfigurability to improve communication performance.
	
	FA technology has also been investigated for physical-layer security. The authors in \cite{ghadi2024FAsecrecy} analyzed the secrecy performance of FA-aided communication systems under spatially correlated fading channels, deriving analytical expressions for average secrecy capacity and secrecy outage probability. The authors in \cite{yao2025fas} jointly optimizes transmit beamforming and FA positions to maximize the secrecy rate under a covertness constraint in the presence of an eavesdropper and a warden. In \cite{li2026FAsecureISAC}, FA positions and transmit beamforming were jointly optimized to maximize the sum secrecy rate in integrated sensing and communication (ISAC) systems. For satellite-terrestrial communications, \cite{li2026FAsatellite} investigated the secrecy outage performance of RIS-assisted FA systems with rate-splitting multiple access under imperfect channel state information. Closely related studies on movable antennas (MAs) have also demonstrated the benefits of antenna position optimization for secure transmission. Specifically, \cite{zhu2023null} investigated array gain enhancement through null steering, \cite{hu2024secure} considered secrecy-rate maximization through joint antenna position and beamforming optimization, and \cite{mei2024discrete} developed a discrete sampling approach for antenna position optimization in secure communications.

	\subsection{Motivation and Contributions}
	
	For satellite communications, existing studies have investigated satellite-mounted MA arrays for dynamic beam coverage and interference suppression \cite{zhu2025dynamic}, energy efficiency maximization \cite{guo2026movable}, and multi-satellite cooperative transmission under statistical channel state information (CSI) \cite{zuo2026joint}. At the GS, MA arrays have been considered for average achievable rate maximization in LEO satellite communications \cite{wang2025joint} and end-to-end sum-rate maximization in dual-hop satellite-terrestrial systems with nonlinear energy harvesting \cite{ali2026dualhop}. Joint beamforming and antenna positioning have also been investigated for LEO ISAC \cite{zhang2025leoisac}, while security-oriented MA designs have been explored in space-air-ground integrated networks (SAGINs) \cite{pang2026secure,wan2026toward}. For FA-aided secure LEO uplinks, three key challenges arise. First, multiple potential eavesdropping satellites may have small angular separations from the serving satellite, limiting the spatial discrimination achievable with a fixed antenna array. Second, orbital motion changes the relative satellite geometry and channel responses over time, requiring adaptive beamforming while avoiding frequent FA reconfiguration. Third, passive eavesdropping satellites do not provide channel feedback, making their CSI difficult to estimate accurately. These challenges motivate the joint optimization of FA positions and time-varying beamforming under imperfect eavesdropping CSI.
	
	In this paper, we investigate FA-aided robust secure uplink transmission in LEO satellite constellations. A GS equipped with a planar FA array transmits confidential information to a serving satellite, while other visible satellites capable of receiving the uplink signal are treated as potential non-colluding eavesdroppers. We assume perfect CSI for the legitimate channel and characterize the uncertainty in the eavesdropping CSI (ECSI) using a norm-bounded error model. To avoid frequent antenna reconfiguration, the FA positions are optimized at the beginning of the observation interval and remain fixed thereafter, whereas transmit beamforming adapts to the time-varying satellite channels. We jointly optimize the FA positions and beamforming to maximize the average worst-case secrecy rate. The main contributions are as follows:
	
	\begin{itemize}
		\item First, we formulate a robust secure uplink transmission problem for an FA-aided LEO satellite constellation. The model accounts for multiple potential eavesdroppers, orbital motion, atmospheric and rain attenuation, Doppler shift, and imperfect ECSI. By jointly optimizing the FA positions and time-varying beamforming, the proposed design exploits FA position reconfigurability to enhance secure transmission while avoiding frequent antenna reconfiguration.
		
		\item Second, we develop a robust AO algorithm to solve the joint design problem. The beamforming subproblem is handled using semidefinite relaxation (SDR) and first-order Taylor approximation, followed by Gaussian randomization for rank-one recovery. The FA position subproblem is solved using successive convex approximation (SCA), which iteratively updates the antenna positions through a sequence of convex optimization problems.
		
		\item Third, we develop a differential evolution (DE)-based AO algorithm to reduce the computational burden of FA position optimization. Instead of repeatedly solving convex surrogate problems, the DE-based method uses population-based search to optimize FA positions while retaining robust beamforming optimization. 
		
		\item Numerical results demonstrate that 1) the proposed FA-aided schemes significantly outperform conventional fixed-position antenna (FPA) schemes in terms of secrecy performance; 2) as the constellation size increases, the optimized FA elements tend to spread toward the boundary of the feasible region, enlarging the effective array aperture and improving spatial discrimination between legitimate and eavesdropping satellites; and 3) the relative performance of the SCA-based and DE-based AO algorithms depends on the number of FA elements. For a small number of FA elements, DE-based AO achieves better performance due to its population-based search capability. As the number of FA elements increases, SCA-based AO can achieve better performance by exploiting the local problem structure, whereas the search efficiency of DE-based AO decreases in the enlarged solution space.
	\end{itemize}

	\textit{Notation:}
	Scalars, vectors, and matrices are denoted by italic letters, boldface lowercase letters, and boldface uppercase letters, respectively. $(\cdot)^{T}$ and $(\cdot)^{H}$ denote the transpose and conjugate transpose, respectively. $|\cdot|$ and $\|\cdot\|_{2}$ denote the absolute value and Euclidean norm, respectively. $\mathrm{Tr}(\cdot)$, $\mathrm{rank}(\cdot)$, and $\mathrm{Re}\{\cdot\}$ denote the trace, rank, and real part, respectively. $\mathbf{I}_{N}$ denotes the $N\times N$ identity matrix, and $\mathbf{A}\succeq\mathbf{0}$ indicates that $\mathbf{A}$ is positive semidefinite. $\lambda_{\max}(\mathbf{A})$ denotes the largest eigenvalue of $\mathbf{A}$. $\nabla_{\mathbf{x}} f$ denotes the gradient of $f$ with respect to $\mathbf{x}$. $\mathcal{A}$ and $|\mathcal{A}|$ denote a set and its cardinality, respectively. Finally, $\jmath=\sqrt{-1}$ denotes the imaginary unit.
	
	\section{SYSTEM MODEL AND PROBLEM FORMULATION}
	\label{pa2}
	\subsection{System Model}
	\begin{figure}[!t]
		\centering
		\includegraphics[width=0.7\columnwidth]{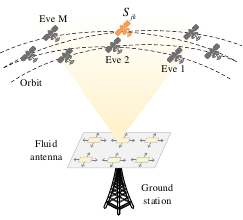}
		\caption{Illustration of FA-aided robust secure uplink transmission in LEO satellite constellations.}
		\label{fig1}
	\end{figure}

\begin{figure}[!t]
		\centering
		\includegraphics[width=0.7\columnwidth]{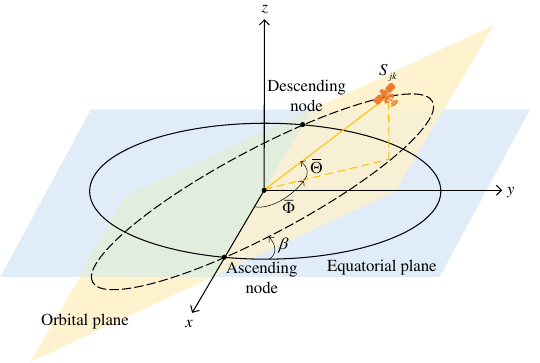}
		\caption{Geometric illustration of a LEO satellite orbit \cite{zhu2025dynamic}.}
		\label{fig2}
	\end{figure}

	As shown in Fig. 1, we consider an uplink communication scenario where a GS equipped with a planar FA array communicates with a LEO satellite constellation. The FA array allows the effective radiating position of each element to be independently configured. The GS transmits the uplink signal to the serving LEO satellite, while the remaining visible satellites on the same orbital shell act as passive eavesdroppers attempting to eavesdrop the signal. A satellite is considered visible to the GS if its elevation angle is no smaller than the minimum elevation threshold of $20^\circ$. We consider the LEO satellite constellation consisting of $J$ equally spaced orbital planes. In each orbital plane, $K$ satellites are initially distributed uniformly along the northbound half of the orbit. Satellites operating in other frequency bands are excluded from the considered constellation. The frequency assignment of each satellite remains unchanged throughout the observation interval. The $k$-th LEO satellite in the $j$-th orbital plane is denoted as ${S_{jk}}$, where $j \in {\cal J} \buildrel \Delta \over = \left\{ {1,2, \cdots ,J} \right\}$, $k \in {\cal K} \buildrel \Delta \over = \left\{ {1,2, \cdots ,K} \right\}$. Fig. \ref{fig2} shows the spatial geometry of a LEO satellite orbit in the geocentric spherical coordinate system (GSCS).

	All the LEO satellites are assumed to move in circular orbits at a common altitude $H$ above the Earth's surface. The position of ${S_{jk}}$ at the continuous time instant $t$ in the GSCS is described by
	\begin{align}
		&\bar R = R + H, ~{\bar \Theta _{jk}}(t) = \arcsin\left[ \sin\beta \,\sin {\alpha _{jk}}(t) \right],\\
		&\bar{\Phi}_{jk}(t)=
		\operatorname{atan2}\left(
		\cos\beta\sin\alpha_{jk}(t),
		\cos\alpha_{jk}(t)
		\right)
		+\frac{2\pi(j-1)}{J},
		\label{eq:satellite_longitude}
	\end{align}
	where $R$ denotes the Earth's radius, $\bar{\Theta}_{jk}(t)\in[-\pi/2,\pi/2]$ and $\bar{\Phi}_{jk}(t)\in(-\pi,\pi]$ denote the geocentric latitude and the geocentric longitude of satellite $S_{jk}$, respectively. $\beta$ denotes the orbital inclination. The geocentric angle of $S_{jk}$ with respect to the ascending node at time \(t\) is expressed as
	${\alpha _{jk}}(t) = {{2\pi t} \mathord{\left/
			{\vphantom {{2\pi t} T}} \right.
			\kern-\nulldelimiterspace} T} + {\alpha _{jk0}}$,
	where ${\alpha _{jk0}} =  - {\pi  \mathord{\left/
			{\vphantom {\pi  2}} \right.
			\kern-\nulldelimiterspace} 2} + {{\pi (k - 1)} \mathord{\left/
			{\vphantom {{\pi (k - 1)} {\left( {K - 1} \right)}}} \right.
			\kern-\nulldelimiterspace} {\left( {K - 1} \right)}}$ denotes the initial angle at $t=0$. 
	The $J$ orbital planes are uniformly spaced in azimuth, and the ascending node of the $j$-th plane is offset by $2\pi (j-1)/J$ from the reference meridian on the equatorial plane. 
	The orbital period is given by $T = 2\pi \sqrt {{{{{\bar R}^3}} \mathord{\left/
				{\vphantom {{{{\bar R}^3}} {{G_e}{M_e}}}} \right.
				\kern-\nulldelimiterspace} {{G_e}{M_e}}}} $, where $G_e$ and $M_e$ represent the gravitational constant and the Earth's mass, respectively. 
	
	Accordingly, the LEO satellite coordinates in the 3D geocentric Cartesian coordinate system (GCCS) can be expressed as
	\begin{equation}
		{\bf R}_{jk}(t) =
		\bar R
		\begin{bmatrix}
			\cos \bar\Theta_{jk}(t)\cos \bar\Phi_{jk}(t)\\
			\cos \bar\Theta_{jk}(t)\sin \bar\Phi_{jk}(t)\\
			\sin \bar\Theta_{jk}(t)
		\end{bmatrix}.
	\end{equation}
	
	The GS is located at a fixed point on the Earth's surface. In the adopted GCCS, its coordinates vary with time due to the Earth's rotation. Thus, the position of the GS is denoted as
	\begin{equation}
		{\bf R}_G(t) =
		R
		\begin{bmatrix}
			\cos\Theta_G\cos\Phi_G(t)\\
			\cos\Theta_G\sin\Phi_G(t)\\
			\sin\Theta_G
		\end{bmatrix},
	\end{equation}
	where $\Theta_G$ denotes the geocentric latitude of the GS, and $\Phi_G(t)=2\pi t/T_E$ represents the time-varying geocentric longitude, with $T_E$ being the Earth's rotation period.
	
	\subsection{Channel Model}
	
	To characterize the array response of the FA array, we further introduce a GS-centered Cartesian coordinate system (SCCS) with its origin located at the GS. The SCCS is defined with its x-, y-, and z-axes pointing toward the local north, east, and radially outward directions, respectively. Let ${\bf{T}}$ denote the transformation matrix from the SCCS to the GCCS. According to the above definition of the axis, ${\bf{T}}$ can be written as
	\begin{align}
		{\bf T}(t) \!=\!\!
		{\small
			\begin{bmatrix}\!
				{ - \sin {\Theta _G}\cos {\Phi _G}(t)} \!&\!{ - \sin {\Phi _G}(t)}\! &\!{\cos {\Theta _G}\cos {\Phi _G}(t)} \\
				{ - \sin {\Theta _G}\sin {\Phi _G}(t)}\! &\!{\cos {\Phi _G}(t)} \!&\!{\cos {\Theta _G}\sin {\Phi _G}(t)} \\
				{\cos {\Theta _G}} \!&0&\!{\sin {\Theta _G}}
		\end{bmatrix}}\!.
	\end{align}
	
	For the uplink transmission, the propagation direction from the GS to satellite $S_{jk}$ in the GCCS is described by
	\begin{equation}
		{\bf d}_{jk}(t) = {\bf R}_{jk}(t) - {\bf R}_G(t).
	\end{equation}
	The corresponding wave vector in the GCCS is expressed as
	\begin{align}
		{{\mathbf{b}}_{jk}}\left( t \right) = \frac{{2\pi {{\bf{d}}_{jk}}\left( t \right)}}{{\lambda {{\left\| {{{\bf{d}}_{jk}}\left( t \right)} \right\|}_2}}},
	\end{align}
	where $\lambda $ denotes the wavelength at the carrier frequency. By applying the coordinate transformation, the wave vector in the SCCS is given by
	\begin{equation}
		{{{\bf{\tilde b}}}_{jk}}\left( t \right) = {\bf{T}}{\left( t \right)^{\rm{T}}}{{\bf{b}}_{jk}}\left( t \right) .
	\end{equation}
	
	The effective radiating position of the $n$-th FA element in the SCCS is given by 
	\begin{equation}
		\bar{\mathbf c}_n
		=
		\big[x_n,\,y_n,\,z_n\big]^{\mathrm T},
	\end{equation}
	where $n \in {\cal N} \buildrel \Delta \over = \left\{ {1, \cdots ,N} \right\}$. Following the planar position model in \cite{liu2026meta}, we constrain the effective radiating positions of the FA elements to the local horizontal plane. Hence, we always have $z_n=0$, and the antenna position can be equivalently expressed as
	\begin{equation}
		\bar{\mathbf c}_n
		=
		\big[{\mathbf c}_n^{\mathrm T},\,0\big]^{\mathrm T},
		\qquad
		{\mathbf c}_n\in\mathcal C\subset\mathbb R^{2},
	\end{equation}
	where $\mathcal C$ denotes the 2D FA reconfiguration region. The radiation characteristics of each FA element are assumed unchanged across different position configurations. Collectively, the FA position vector is defined as
	\begin{equation}
		{\mathbf c}
		=
		\big[
		{\mathbf c}_1^{\mathrm T},
		{\mathbf c}_2^{\mathrm T},
		\ldots,
		{\mathbf c}_N^{\mathrm T}
		\big]^{\mathrm T},
	\end{equation}
	which determines the geometry of the FA array. Hence, the transmit steering vector of the FA array towards satellite ${S_{jk}}$ at time $t$ is given as
	\begin{align}
		{{\bf{s}}_{jk}}({{\bf{\tilde b}}_{jk}}\left( t \right),{\bf{c}}) = {[{e^{j{\mkern 1mu} {{{\bf{\tilde b}}}_{jk}}{{\left( t \right)}^{\rm{T}}}{{\overline {\bf{c}} }_1}}}, \ldots ,{e^{j{\mkern 1mu} {{{\bf{\tilde b}}}_{jk}}{{\left( t \right)}^{\rm{T}}}{{\overline {\bf{c}} }_N}}}]^{\rm{T}}}.
	\end{align}
	
	For the channel from the GS to satellite ${S_{jk}}$, the corresponding large-scale channel gain is modeled as
	\begin{equation}
		\rho_{jk}(t)
		=
		\rho_0
		\left\|\mathbf d_{jk}(t)\right\|_2^{-\alpha}
		D\!\left(\tau_{jk}(t)\right)
		L_{jk}^{\rm atm}(t)
		L_{jk}^{\rm rain}(t),
		\label{eq:large_scale_gain}
	\end{equation}
	where $\rho_0$ denotes the path gain at the reference distance $d_0=1$ m, and $\alpha$ is the path-loss exponent. 
	$D\!\left(\tau_{jk}(t)\right)$ denotes the directional receive antenna gain of satellite $S_{jk}$, where $\tau_{jk}(t)$ represents the off-boresight angle between the satellite antenna boresight and the satellite-to-GS direction. The directional receive antenna gain of satellite $S_{jk}$ is modeled following \cite{3gpp2020study} as
	\begin{equation}
		D\bigl(\tau_{jk}(t)\bigr)
		=D_0\left[\frac{2J_1(\xi_{jk}(t))}
		{\xi_{jk}(t)}\right]^2,
		\label{eq:antenna_gain}
	\end{equation}
	where $\xi_{jk}(t)=2\pi r\sin\tau_{jk}(t)/\lambda$,
	$J_1(\cdot)$ is the first-order Bessel function of the first kind,
	and $D_0=\eta(2\pi r/\lambda)^2$ denotes the maximum antenna gain,
	with $r$ and $\eta$ representing the satellite antenna aperture radius
	and efficiency, respectively. At $\xi_{jk}(t)=0$, the normalized
	gain is defined as unity by continuity. Moreover, the atmospheric and rain attenuation factors are respectively expressed as
	\begin{equation}
		L_{jk}^{\rm atm}(t)
		=
		10^{-A_{jk}^{\rm atm}(t)/10},
		\
		L_{jk}^{\rm rain}(t)
		=
		10^{-A_{jk}^{\rm rain}(t)/10},
		\label{eq:attenuation_factors}
	\end{equation}
	where $A_{jk}^{\rm atm}(t)$ and $A_{jk}^{\rm rain}(t)$ denote the atmospheric attenuation and rain attenuation in dB, respectively. Due to the high mobility of LEO satellites, the time-varying propagation distance induces a Doppler shift. Let $d_{jk}(t)=\|\mathbf d_{jk}(t)\|_2$ denote the distance between satellite $S_{jk}$ and the GS. The instantaneous Doppler shift is given by $f_{D,jk}(t)=-\dot d_{jk}(t)/\lambda$, and the corresponding Doppler-induced phase is $\phi_{D,jk}(t)=2\pi\int_{0}^{t}f_{D,jk}(\tau)\,{\rm d}\tau$. Accordingly, the uplink channel between the FA array and satellite $S_{jk}$ is modeled as
	\begin{equation}
		\hat{\mathbf h}_{jk}(\mathbf c,t)
		=
		\sqrt{\rho_{jk}(t)}
		e^{j\left(\phi_{0,jk}+\phi_{D,jk}(t)\right)}
		\mathbf s_{jk}\!\left(\tilde{\mathbf b}_{jk}(t),\mathbf c\right),
		\label{eq:channel_model}
	\end{equation}
	where $\phi_{0,jk}$ denotes the initial propagation phase.
	
	Among all visible satellites, the GS selects the satellite with the largest large-scale channel gain as the legitimate receiver, i.e., $S_L=S_{j_0k_0}$ with $(j_0,k_0)=\arg\max_{(j,k)\in\mathcal V(t)}\rho_{jk}(t)$, where $\mathcal V(t)$ denotes the visible-satellite set at time $t$. Since $S_L$ cooperates with the GS for channel acquisition, its CSI is assumed to be perfectly known. Thus,
	\begin{align}
		{\bf h}_L({\bf c},t)
		&=\hat{\bf h}_{j_0k_0}({\bf c},t),\\
		y_L({\bf c},{\bf w}(t),t)
		&={\bf w}^{\rm H}(t){\bf h}_L({\bf c},t)x(t)+n_L(t).
	\end{align}
	The remaining visible satellites are conservatively regarded as potential passive eavesdroppers. Denote their set by $\mathcal E(t)=\mathcal V(t)\setminus\{S_L\}$ and let $\mathcal M(t)\triangleq\{1,\ldots,M(t)\}$ index the $M(t)=|\mathcal E(t)|$ eavesdroppers. For the $m$-th eavesdropper $S_{E,m}$, only nominal ECSI is assumed to be available, and its actual channel is modeled as
	\begin{equation}
		\begin{aligned}
			&{\bf h}_{E,m}({\bf c},t)
			=
			\hat{\bf h}_{E,m}({\bf c},t)
			+
			\Delta{\bf h}_{E,m}(t),\\
			&\|\Delta{\bf h}_{E,m}(t)\|_2 \le \varepsilon_{E,m}(t).
		\end{aligned}
	\end{equation}
	The received signal at $S_{E,m}$ can be written as\footnotemark
	\footnotetext{In each time slot, the GS communicates with only one authorized LEO satellite, while all the other visible satellites are treated as potential eavesdroppers. Due to the high mobility of LEO satellites, the serving satellite may change over time to maintain service continuity.}
	\begin{align}
		y_{E,m}({\bf c},{\bf w}(t),t)
		=
		{\bf w}^{\rm H}(t){\bf h}_{E,m}({\bf c},t)x(t)
		+n_{E,m}(t),
	\end{align}
	where $n_L(t)$ and $n_{E,m}(t)$ denote the additive noises at the legitimate satellite and the $m$-th eavesdropper, respectively. The received signal-to-noise ratio (SNR) at the legitimate satellite is given by
	\begin{align}
		{\gamma _L}({\bf{c}},{\bf{w}}(t),t) = \frac{{|{{\bf{w}}^{\rm{H}}}(t){{\bf{h}}_L}({\bf{c}},t){|^2}}}{{\sigma _L^2}}.
	\end{align}
	For the $m$-th eavesdropper, considering the norm-bounded ECSI uncertainty \cite{chen2022robust}, its worst-case SNR is defined as
	\begin{equation}
		\begin{aligned}
			&\gamma_{E,m}^{\rm wc}({\bf c},{\bf w}(t),t)=\\
			&\max_{\|\Delta{\bf h}_{E,m}(t)\|_2\leq\varepsilon_{E,m}(t)}
			\frac{\Big|
				{\bf w}^{\rm H}(t)
				\big(
				\hat{\bf h}_{E,m}({\bf c},t)+\Delta{\bf h}_{E,m}(t)
				\big)
				\Big|^2}{\sigma_{E,m}^2}.
		\end{aligned}
	\end{equation}
	By the Cauchy--Schwarz inequality, the above worst-case SNR can be equivalently expressed as
	\begin{equation}
		\gamma _{E,m}^{{\rm{wc}}}({\bf{c}},{\bf{w}}(t),t) = \frac{{{{\left(\left| {{{\bf{w}}^{\rm{H}}}(t){{\widehat {\bf{h}}}_{E,m}}({\bf{c}},t)} \right| + {\varepsilon _{E,m}}(t){{\left\| {{\bf{w}}(t)} \right\|}_2}\right)}^2}}}{{\sigma _{E,m}^2}}.
	\end{equation}
	Accordingly, the instantaneous worst-case secrecy rate is
	\begin{equation}
		\begin{aligned}
			&C_s^{\rm wc}({\bf c},{\bf w}(t),t)
			= \Big[\log_2(1+\gamma_L({\bf c},{\bf w}(t),t))\\
			&~~~~~~~~~-\max_{m\in{\cal M}(t)}
			\log_2(1+\gamma_{E,m}^{\rm wc}({\bf c},{\bf w}(t),t))
			\Big]^+.
		\end{aligned}
	\end{equation}
	where $\sigma_L^2$ and $\sigma_{E,m}^2$ denote the noise powers at $S_L$ and $S_{E,m}$, respectively, and $[x]^+\triangleq\max\{x,0\}$.
	
	\subsection{Problem Formulation}
	
	Satellites in the same orbital plane follow the same trajectory with different initial phase offsets. Due to satellite orbital motion and the Earth's rotation, the relative geometry between the GS and the constellation varies over time. To characterize the time-varying satellite geometry, we set the observation period to $\bar T=T_E/J$, during which the Earth rotates by an angle of $2\pi/J$, equal to the angular spacing between neighboring orbital planes. The interval $(0,\bar T]$ is uniformly divided into $P$ time slots, with the midpoint of the $p$-th slot given by $t_p=(p-\frac{1}{2})\bar T/P$, $p\in\mathcal P\triangleq\{1,\ldots,P\}$. In each time slot, the serving satellite and potential eavesdroppers are determined based on the instantaneous satellite geometry. To avoid frequent antenna reconfiguration, the FA position vector ${\bf c}$ is optimized once for the observation interval and remains fixed throughout it, while the transmit beamforming vector ${\bf w}[p]$ is dynamically adjusted in each time slot.
	
	Let $\mathcal M[p]\triangleq\mathcal M(t_p)$ denote the index set of the potential eavesdroppers visible in the $p$-th slot. The legitimate and worst-case eavesdropping SNRs are respectively given by
	\begin{align}
		\gamma_L({\bf c},{\bf w}[p],t_p)
		=
		\frac{\left|{\bf w}^{\rm H}[p]{\bf h}_L({\bf c},t_p)\right|^2}
		{\sigma_L^2},
	\end{align}
	\begin{align}
		\gamma _{E,m}^{{\rm{wc}}}(\!{\bf{c}},{\bf{w}}[p],{t_p}) \!= \!\frac{{{{(\left| {{{\bf{w}}^{\rm{H}}}[p]{{\widehat {\bf{h}}}_{E,m}}({\bf{c}},{t_p})} \right| \!+ \!{\varepsilon _{E,m}}(\!{t_p}){{\left\| \!{{\bf{w}}[p]} \right\|}_2})}^2}}}{{{\sigma _{E,m}}\!{^2}}},
		\label{p1p2}
	\end{align}
	where $m\in\mathcal M[p]$. The corresponding achievable rates are
	\begin{align}
		A_L({\bf c},{\bf w}[p],t_p)=\log_2\!\left(1+\gamma_L({\bf c},{\bf w}[p],t_p)\right),
	\end{align}
	\begin{align}
		A_{E,m}^{\rm wc}({\bf c},{\bf w}[p],t_p)=\log_2\left(1+\gamma_{E,m}^{\rm wc}({\bf c},{\bf w}[p],t_p)\right).
	\end{align}
	Accordingly, the worst-case secrecy rate in the $p$-th slot is
	\begin{equation}
		\begin{aligned}
			C_s^{\rm wc}({\bf c},{\bf w}[p],t_p)
			=
			\Big[
			&A_L({\bf c},{\bf w}[p],t_p)\\
			&-\max_{m\in\mathcal M[p]}
			A_{E,m}^{\rm wc}({\bf c},{\bf w}[p],t_p)
			\Big]^+.
		\end{aligned}
	\end{equation}
	The average worst-case secrecy rate over $(0,\bar T]$ is approximated as
	\begin{equation}
		\bar C_s^{\rm wc}\left({\bf c},\{{\bf w}[p]\}_{p=1}^{P}\right)
		=
		\frac{1}{P}\sum_{p=1}^{P}
		C_s^{\rm wc}({\bf c},{\bf w}[p],t_p).
	\end{equation}
	
	The objective is to maximize the average worst-case secrecy rate by jointly optimizing the transmit beamforming vectors and the FA position vector, subject to the minimum secrecy rate, transmit power, FA reconfiguration region, and minimum inter-antenna spacing constraints. Accordingly, the optimization problem is formulated as
	\begin{subequations}
		\begin{align}
			({\rm P1}):\quad 
			&\mathop{\max}\limits_{{\bf c},\left\{{\bf w}[p]\right\}_{p=1}^P}
			\ \bar C_s^{\rm wc}\!\left({\bf c},\left\{{\bf w}[p]\right\}_{p=1}^P\right) \\
			&\text{s.t.}\quad
			C_s^{\rm wc}\!\left({\bf c},{\bf w}[p],t_p\right)
			\ge C_{\min},\quad \forall p\in\mathcal P, \label{p14}\\
			&\left\|{\bf w}[p]\right\|_2^2
			\le P_{\max},\quad \forall p\in\mathcal P, \label{p18}\\
			&{\bf c}_n\in\mathcal C,\quad \forall n\in\mathcal N, \label{p3}\\
			&\left\|{\bf c}_n-{\bf c}_{\tilde n}\right\|_2
			\ge d_{\min}, \forall n,\tilde n\in\mathcal N,\; n\ne\tilde n. \label{p4}
		\end{align}
	\end{subequations}
	where $C_{\min}$ denotes the minimum required secrecy rate, $P_{\max}$ is the maximum transmit power budget, and $d_{\min}$ represents the minimum allowable distance between any two antenna elements. Constraint (\ref{p14}) guarantees the minimum worst-case secrecy rate in each time slot, constraint (\ref{p18}) limits the GS transmit power, constraint (\ref{p3}) specifies the feasible reconfiguration region of each FA element, and constraint (\ref{p4}) enforces the minimum inter-antenna spacing. Problem (P1) is non-convex because the FA positions and transmit beamforming vectors are coupled in the secrecy rate expressions, the worst-case secrecy-rate constraint (\ref{p14}) is non-convex, and the spacing constraint (\ref{p4}) defines a non-convex feasible set.
	
	\section{Proposed SCA-Based AO Algorithm}
	\label{pa3}
	
	To address the above problem, an AO method is proposed, where the FA position and beamforming vectors are updated alternately. Specifically, the $\{{\bf w}[p]\}_{p=1}^P$ are optimized with $\bf c$ fixed, while $\bf c$ is optimized with fixed $\{{\bf w}[p]\}_{p=1}^P$.
	
	\subsection{Optimization of $\{{\bf w}[p]\}_{p=1}^{P}$ with Fixed ${\bf c}$}
	
	With the FA position vector ${\bf c}$ fixed, define $A_E^{\rm wc}({\bf w}[p],t_p)=\max_{m\in\mathcal M[p]}A_{E,m}^{\rm wc}({\bf w}[p],t_p)$. Introducing auxiliary variables $\tau[p]\geq0$ and $r_E[p]\geq0$, we have
	\begin{equation}
		\tau[p]\leq A_L({\bf w}[p],t_p)-r_E[p],
	\end{equation}
	\begin{equation}
		A_{E,m}^{\rm wc}({\bf w}[p],t_p)\leq r_E[p].
	\end{equation}
	Let ${\bf W}[p]={\bf w}[p]{\bf w}^{\rm H}[p]$ and ${\bf H}_L[p]={\bf h}_L(t_p){\bf h}_L^{\rm H}(t_p)$.
	Moreover, introduce $q_{E,m}[p]\geq0$ to upper-bound the worst-case received signal power at the $m$-th eavesdropper, we define
	\begin{equation}
		\begin{aligned}
			&\max_{\Delta{\bf h}_{E,m}(t_p)\in\mathcal U_{E,m}(t_p)}
			\big(\hat{\bf h}_{E,m}(t_p)
			+\Delta{\bf h}_{E,m}(t_p)\big)^{\rm H}
			{\bf W}[p] \\
			&\times
			\big(\hat{\bf h}_{E,m}(t_p)
			+\Delta{\bf h}_{E,m}(t_p)\big)
			\leq q_{E,m}[p].
		\end{aligned}
		\label{robust_power}
	\end{equation}
	By applying the S-procedure to constraint (\ref{robust_power}), it can be equivalently transformed into the linear matrix inequality (LMI) in (\ref{Sprocedure}), where $\lambda_{E,m}[p]\geq 0$ is an auxiliary scalar introduced by the S-procedure. For clarity, the corresponding LMI is presented at the top of this page.
	
	\begin{figure*}[!t]
		
		\vspace{4pt}
		\begin{equation}
			{\bf \Psi}_{E,m}[p]\triangleq
			\begin{bmatrix}
				\lambda_{E,m}[p]{\bf I}-{\bf W}[p]
				&
				-{\bf W}[p]\hat{\bf h}_{E,m}(t_p)
				\\
				-\hat{\bf h}_{E,m}^{\rm H}(t_p){\bf W}[p]
				&
				q_{E,m}[p]
				-\hat{\bf h}_{E,m}^{\rm H}(t_p)
				{\bf W}[p]
				\hat{\bf h}_{E,m}(t_p)
				-\lambda_{E,m}[p]\varepsilon_{E,m}^2(t_p)
			\end{bmatrix}
			\succeq{\bf 0}.
			\label{Sprocedure}
		\end{equation}
		\hrule
	\end{figure*}
	
	Accordingly, the beamforming subproblem can be reformulated as
	\begin{subequations}
		\begin{align}
			({\rm P2}):
			&\max_{\substack{
					{\bf W}[p],\tau[p],r_E[p],\\
					q_{E,m}[p],\lambda_{E,m}[p]}}
			\frac{1}{P}\sum_{p=1}^{P}\tau[p] \label{P2a}\\
			&{\rm s.t.}\quad\tau[p]\geq C_{\min}, \forall p, \label{P2b}\\
			&{\log _2}(1 + \frac{{{\rm{Tr}}({{\bf{H}}_L}[p]{\bf{W}}[p])}}{{\sigma _L^2}}) \ge \tau [p] + {r_E}[p],\forall p, \label{P2c}\\
			&{\log _2}(1 + \frac{{{q_{E,m}}[p]}}{{\sigma _{E,m}^2}}) \le {r_E}[p],\forall m \in {\cal M}[p],{\mkern 1mu} p, \label{P2d}\\
			&{\bf \Psi}_{E,m}[p]\succeq{\bf 0},\lambda_{E,m}[p]\geq0,\forall m\in\mathcal M[p],\,p, \label{P2e}\\
			&{\rm Tr}({\bf W}[p])\leq P_{\max},\label{P2f}\\
			&{\bf W}[p]\succeq{\bf 0},\forall p, \label{P2g}\\
			&{\rm rank}({\bf W}[p])=1,\forall p. \label{P2h}
		\end{align}
	\end{subequations}
	By applying SDR, the rank-one constraint (\ref{P2h}) is first relaxed. However, the resulting problem remains non-convex due to constraint (\ref{P2d}). Define
	$f_{E,m}(q_{E,m}[p])=\log_2(1+q_{E,m}[p]/\sigma_{E,m}^2)$. Since $f_{E,m}(q_{E,m}[p])$ is concave, its first-order Taylor expansion at $q_{E,m}^{(l)}[p]$ provides a global upper bound
	\begin{equation}
		\begin{aligned}
			{f_{E,m}}({q_{E,m}}[p]) \le {f_{E,m}}(q_{E,m}^{(l)}[p]) + \frac{{{q_{E,m}}[p] - q_{E,m}^{(l)}[p]}}{{\ln 2(\sigma _{E,m}^2 + q_{E,m}^{(l)}[p])}}.
		\end{aligned}
		\label{Eve_upper_bound}
	\end{equation}
	Let $\bar f_{E,m}^{(l)}(q_{E,m}[p])$ denote the right-hand side of (\ref{Eve_upper_bound}). By imposing the convex constraint $\bar f_{E,m}^{(l)}(q_{E,m}[p])\leq r_E[p]$ in place of (\ref{P2d}) and dropping the rank-one constraint (\ref{P2h}), we obtain the following convex approximation of (P2)
	\begin{subequations}
		\begin{align}
			({\rm P3}):
			&\max_{\substack{
					{\bf W}[p],\tau[p],r_E[p],\\
					q_{E,m}[p],\lambda_{E,m}[p]}}
			\frac{1}{P}\sum_{p=1}^{P}\tau[p] \\
			&{\rm s.t.}\quad
			(\ref{P2b}),(\ref{P2c}),(\ref{P2e}),(\ref{P2f}),(\ref{P2g}),\\
			&\bar f_{E,m}^{(l)}(q_{E,m}[p])\leq r_E[p],\forall m\in\mathcal M[p],\;p\in\mathcal P.
		\end{align}
	\end{subequations}
	
	Problem (P3) can be efficiently solved \cite{10702481}. When ${\bf W}^{\star}[p]$ is rank one, the beamforming vector can be obtained directly from its dominant eigenvalue and corresponding eigenvector. Otherwise, Gaussian randomization is employed to obtain a feasible rank-one beamforming solution. Specifically, the relaxed solution is decomposed as
	\begin{align}
		{\bf W}^{\star}[p]
		=
		{\bf U}_p{\bf \Sigma}_p{\bf U}_p^{\rm H},
	\end{align}
	where ${\bf U}_p\in\mathbb C^{N\times N}$ and ${\bf \Sigma}_p\in\mathbb C^{N\times N}$ denote the eigenvector matrix and the diagonal eigenvalue matrix, respectively. For each randomization trial, a candidate beamforming vector is generated as
	\begin{align}
		\tilde{\bf w}[p]
		=
		\sqrt{P_{\max}}
		\frac{
			{\bf U}_p{\bf \Sigma}_p^{1/2}{\bf r}
		}{
			\left\|
			{\bf U}_p{\bf \Sigma}_p^{1/2}{\bf r}
			\right\|_2
		},
	\end{align}
	where ${\bf r}\sim\mathcal{CN}({\bf 0},{\bf I}_N)$. The worst-case secrecy rate of each candidate is then evaluated using the bounded-ECSI model, and candidates violating the minimum secrecy rate constraint are discarded. Among the feasible candidates, the one achieving the largest average worst-case secrecy rate is selected as the beamforming solution.
	
	\subsection{Optimization of ${\bf{c}}$ with Fixed ${\left\{ {{\bf{w}}\left[ p \right]} \right\}_{p = 1}^P}$}
	With $\{{\bf w}[p]\}_{p=1}^{P}$ fixed, (P1) can be reformulated as
	\begin{subequations}
		\begin{align}
			({\rm P4}):\quad
			&\max_{\bf c}\ 
			\bar C_s^{\rm wc}\!\left(
			{\bf c}\right)\\
			&{\rm s.t.}\quad
			(\ref{p14}),\;(\ref{p3}),\;(\ref{p4}).
		\end{align}
	\end{subequations}
	Problem (P4) remains non-convex due to the worst-case secrecy rate constraint (\ref{p14}) and the minimum inter-antenna spacing constraint (\ref{p4}). Therefore, we adopt an SCA framework, in which tractable surrogate functions are iteratively constructed around the current FA position vector.
	
	First, to handle the non-smooth maximum operation over the potential eavesdroppers, we adopt the log-sum-exp function as a smooth approximation. Specifically, the worst-case eavesdropping rate in the $p$-th time slot is approximated by
	\begin{equation}
		\tilde A_E^{{\rm{wc}}}({\bf{c}},{t_p}) = \mu \ln (\sum\limits_{m \in M[p]} {{e^{{{A_{E,m}^{{\rm{wc}}}({\bf{c}},{t_p})} \mathord{\left/{\vphantom {{A_{E,m}^{{\rm{wc}}}({\bf{c}},{t_p})} \mu }} \right.
							\kern-\nulldelimiterspace} \mu }}}} ),
		\label{smooth_eve}
	\end{equation}
	where $\mu>0$ is the smoothing parameter. As $\mu\rightarrow0$, $\tilde A_E^{\rm wc}({\bf c},t_p)$ approaches $\max_{m\in\mathcal M[p]}A_{E,m}^{\rm wc}({\bf c},t_p)$. Accordingly, the smoothed average worst-case secrecy rate is given by
	\begin{equation}
		\tilde C_s^{\rm wc}({\bf c})
		=
		\frac{1}{P}\sum_{p=1}^{P}
		\left[
		A_L({\bf c},t_p)
		-
		\tilde A_E^{\rm wc}({\bf c},t_p)
		\right].
		\label{smooth_secrecy}
	\end{equation}
	We next derive the gradient of (\ref{smooth_secrecy}) with respect to the FA position vector. For the legitimate satellite, define
	\begin{equation}
		u_L({\bf c},t_p)
		=
		{\bf s}_L^{\rm H}(\tilde{\bf b}_L(t_p),{\bf c})
		{\bf W}[p]
		{\bf s}_L(\tilde{\bf b}_L(t_p),{\bf c}),
	\end{equation}
	\begin{equation}
		a_L(t_p)=\rho_L(t_p)/\sigma_L^2,
	\end{equation}
	where ${\bf W}[p]={\bf w}[p]{\bf w}^{\rm H}[p]$. Thus,
	\begin{equation}
		\gamma_L({\bf c},t_p)=a_L(t_p)u_L({\bf c},t_p). 
	\end{equation}
	For the $m$-th eavesdropper, similarly define
	\begin{equation}
		u_{E,m}({\bf c},t_p)
		=
		{\bf s}_{E,m}^{\rm H}(\tilde{\bf b}_{E,m}(t_p),{\bf c})
		{\bf W}[p]
		{\bf s}_{E,m}(\tilde{\bf b}_{E,m}(t_p),{\bf c}).
	\end{equation}
	According to the norm-bounded ECSI model, the worst-case eavesdropping SNR can be rewritten as
	\begin{equation}
		\gamma _{E,m}^{{\rm{wc}}}({\bf{c}},{t_p}) = \frac{{{{(\sqrt {{\rho _{E,m}}({t_p}){u_{E,m}}({\bf{c}},{t_p})}  + {\varepsilon _{E,m}}({t_p})\|{\bf w}[p]\|_2)}^2}}}{{\sigma _{E,m}^2}}.
		\label{robust_snr_position}
	\end{equation}
	
	For $\chi\in\{L,(E,m)\}$, the $n$-th entry of the corresponding steering vector is $s_{\chi,n}=e^{j{\bf k}_{\chi}^{\rm T}(t_p){\bf c}_n}$, where ${\bf k}_{\chi}(t_p)$ denotes the horizontal component of $\tilde{\bf b}_{\chi}(t_p)$. Hence, $\partial s_{\chi,n}/\partial{\bf c}_n
	=js_{\chi,n}{\bf k}_{\chi}(t_p)$. It follows that
	\begin{equation}
		\frac{\partial u_{\chi}({\bf c},t_p)}
		{\partial{\bf c}_n}
		=
		2\,{\rm Im}\!\left\{
		s_{\chi,n}^{\rm H}
		\big({\bf W}[p]{\bf s}_{\chi}\big)_n
		\right\}
		{\bf k}_{\chi}(t_p).
		\label{p15}
	\end{equation}
	By the chain rule, the gradient of the legitimate achievable rate is
	\begin{equation}
		\frac{\partial A_L({\bf c},t_p)}
		{\partial{\bf c}_n}
		=
		\frac{
			a_L(t_p)
		}{
			\ln2\left(1+a_L(t_p)u_L({\bf c},t_p)\right)
		}
		\frac{\partial u_L({\bf c},t_p)}
		{\partial{\bf c}_n}.
		\label{grad_legitimate}
	\end{equation}
	For notational convenience, define $\xi_{E,m}({\bf c},t_p) = \sqrt{\rho_{E,m}(t_p)u_{E,m}({\bf c},t_p)}
	+ \varepsilon_{E,m}(t_p)\|{\bf w}[p]\|_2$. Then, the gradient of the worst-case eavesdropping rate is
	\begin{equation}
		\begin{aligned}
			\frac{\partial A_{E,m}^{\rm wc}({\bf c},t_p)}
			{\partial{\bf c}_n}
			=&
			\frac{
				\sqrt{\rho_{E,m}(t_p)}
				\,\xi_{E,m}({\bf c},t_p)
			}{
				\ln2\,\sigma_{E,m}^2
				\sqrt{u_{E,m}({\bf c},t_p)}
			}\\
			&\times
			\frac{
				1
			}{
				1+\gamma_{E,m}^{\rm wc}({\bf c},t_p)
			}
			\frac{\partial u_{E,m}({\bf c},t_p)}
			{\partial{\bf c}_n}.
		\end{aligned}
		\label{grad_eve}
	\end{equation}
	
	Using the chain rule, the gradient of the smoothed worst-case eavesdropping rate is
	\begin{equation}
		\nabla_{\bf c}\tilde A_E^{\rm wc}
		=
		\sum_{m\in\mathcal M[p]}
		\omega_{E,m}[p]\,
		\nabla_{\bf c}A_{E,m}^{\rm wc},
		\label{p22}
	\end{equation}
	where
	\begin{equation}
		{\omega _{E,m}}[p] = \frac{{{e^{(A_{E,m}^{{\rm{wc}}}({\bf{c}},{t_p})/\mu )}}}}{{\sum\limits_{i \in M[p]} {{e^{^{(A_{E,i}^{{\rm{wc}}}({\bf{c}},{t_p})/\mu )}}}} }}.
	\end{equation}
	Accordingly,
	\begin{equation}
		{\nabla _{\bf{c}}}\tilde C_s^{{\rm{wc}}} = \frac{1}{P}\sum\limits_{p = 1}^P {({\nabla _{\bf{c}}}{A_L} - {\nabla _{\bf{c}}}\tilde A_E^{{\rm{wc}}})} .
		\label{grad_secrecy}
	\end{equation}
	According to (\ref{p15}), (\ref{grad_legitimate}), (\ref{grad_eve}), and (\ref{p22}), the complete gradient of the smoothed average worst-case secrecy rate is given in (\ref{grad_full}).
	\begin{figure*}[!b]
		\hrule
		\begin{equation}
			\begin{aligned}
				\nabla_{{\bf c}_n}\tilde C_s^{\rm wc}({\bf c})
				=
				&\frac{1}{P}\sum_{p=1}^{P}
				\Bigg[
				\frac{
					2a_L(t_p)
				}{
					\ln2\left(1+a_L(t_p)u_L({\bf c},t_p)\right)
				}
				{\rm Im}\!\left\{
				s_{L,n}^{\rm H}
				\big({\bf W}[p]{\bf s}_L\big)_n
				\right\}
				{\bf k}_L(t_p)
				\\
				&-
				\sum_{m\in\mathcal M[p]}
				\omega_{E,m}[p]\,
				\frac{
					2\sqrt{\rho_{E,m}(t_p)}
					\,\xi_{E,m}({\bf c},t_p)
				}{
					\ln2\,\sigma_{E,m}^2
					\sqrt{u_{E,m}({\bf c},t_p)}
					\left(
					1+\gamma_{E,m}^{\rm wc}({\bf c},t_p)
					\right)
				}
				{\rm Im}\!\left\{
				s_{E,m,n}^{\rm H}
				\big({\bf W}[p]{\bf s}_{E,m}\big)_n
				\right\}
				{\bf k}_{E,m}(t_p)
				\Bigg].
			\end{aligned}
			\label{grad_full}
		\end{equation}
	\end{figure*}
	Since the smoothed secrecy-rate function in (\ref{smooth_secrecy}) is generally non-concave with respect to ${\bf c}$, its first-order Taylor expansion alone does not necessarily constitute a global lower bound. Therefore, at the current iterate ${\bf c}^{(r)}$, we construct the following concave quadratic surrogate
	\begin{equation}
		\begin{aligned}
			\underline C_s^{(r)}({\bf c})
			=&\,
			\tilde C_s^{\rm wc}({\bf c}^{(r)})+\nabla_{\bf c}\tilde C_s^{\rm wc}({\bf c}^{(r)})^{\rm T}
			({\bf c}-{\bf c}^{(r)})\\
			&-\frac{\zeta^{(r)}}{2}\|{\bf c}-{\bf c}^{(r)}\|_2^2,
		\end{aligned}
		\label{secrecy_surrogate}
	\end{equation}
	where $\zeta^{(r)}>0$ is chosen sufficiently large such that $\underline C_s^{(r)}({\bf c})$ provides a locally tight lower surrogate of the smoothed secrecy-rate function.
	
	The minimum inter-antenna spacing constraint (\ref{p4}) is also non-convex. Since the Euclidean norm is convex, its first-order Taylor expansion at the current point ${\bf c}^{(r)}$ provides a global lower bound. Hence, constraint (\ref{p4}) can be conservatively approximated as
	\begin{equation}
		\frac{{{{({\bf{c}}_n^{(r)} - {\bf{c}}_{\tilde n}^{(r)})}^{\rm{T}}}}}{{{{\left\| {{\bf{c}}_n^{(r)} - {\bf{c}}_{\tilde n}^{(r)}} \right\|}_2}}}{{\bf{c}}_n} - {{\bf{c}}_{\tilde n}} \ge {d_{\min }},\;\;\;{\mkern 1mu} \forall n \ne \tilde n.
		\label{spacing_surrogate}
	\end{equation}
	For the per-slot secrecy-rate constraint, define
	\begin{equation}
		F_p({\bf c})
		=
		A_L({\bf c},t_p)
		-
		\tilde A_E^{\rm wc}({\bf c},t_p).
	\end{equation}
	At the current point ${\bf c}^{(r)}$, a locally tight lower surrogate of $F_p({\bf c})$ is constructed as
	\begin{equation}
		\begin{aligned}
			\underline F_p^{(r)}({\bf c})
			\!=\!
			F_p({\bf c}^{(r)})
			\!+\!
			\nabla_{\bf c}F_p({\bf c}^{(r)})^{\rm T}
			({\bf c}\!-\!{\bf c}^{(r)})\!-\!
			\frac{\zeta_p^{(r)}}{2}
			\|{\bf c}\!-\!{\bf c}^{(r)}\|_2^2,
		\end{aligned}
		\label{slot_surrogate}
	\end{equation}
	where $\zeta_p^{(r)}>0$ is chosen to ensure the required lower-bound property. Therefore, the minimum secrecy-rate constraint can be approximated by
	\begin{equation}
		\underline F_p^{(r)}({\bf c})
		\ge C_{\min},
		\quad
		\forall p\in\mathcal P.
		\label{secrecy_constraint_surrogate}
	\end{equation}
	
	At the $r$-th SCA iteration, the FA position subproblem is reformulated using the derived surrogate functions as
	\begin{subequations}
		\begin{align}
			({\rm P5}):\quad
			&\max_{\bf c}\ \underline C_s^{(r)}({\bf c})\\
			&{\rm s.t.}\quad (\ref{spacing_surrogate}),(\ref{secrecy_constraint_surrogate}), (\ref{p3}).
		\end{align}
	\end{subequations}
	Problem (P5) is convex and can be efficiently solved using CVX. The curvature parameters $\zeta^{(r)}$ and $\zeta_p^{(r)}$ can be adjusted via backtracking to ensure the lower-bound conditions of the surrogate functions.

	\subsection{Convergence and Computational Complexity}
	We propose an AO algorithm to alternately update the beamforming vectors $\{{\bf w}[p]\}_{p=1}^{P}$ and the FA position vector ${\bf c}$. The beamforming subproblem is addressed using SDR and first-order approximation, with Gaussian randomization applied when rank-one recovery is required. A candidate solution is retained only if it satisfies the secrecy-rate constraints and achieves an average worst-case secrecy rate no lower than the previous solution. Similarly, the FA position subproblem is solved via SCA, and the resulting update is accepted only if it preserves feasibility and does not reduce the objective value. If either update fails to meet these criteria, the previous feasible solution is retained. Thus, given a feasible initialization, the proposed algorithm generates a nondecreasing sequence of objective values, up to numerical tolerances. Since the objective is bounded above under the finite transmit-power constraint, the objective sequence converges in the idealized zero-tolerance setting. The overall procedure is summarized in {\bf Algorithm 1}.
	
	\begin{algorithm}[!t]
		\caption{Robust SCA-Based AO Algorithm for Solving (P1)}
		\label{alg:AO_one}
		\begin{algorithmic}[1]
			\STATE \textbf{Initialize:} Set $r=0$, and choose feasible $\mathbf c^{(0)}$
			and $\{\mathbf w^{(0)}[p]\}_{p=1}^{P}$.
			\REPEAT
			\STATE Given $\mathbf c^{(r)}$, initialize
			$\{q_{E,m}^{(0)}[p]\}$ and set $l=0$.
			\REPEAT
			\STATE Construct the upper bounds in (\ref{Eve_upper_bound}) at
			$\{q_{E,m}^{(l)}[p]\}$ and solve (P3) via CVX.
			\STATE Update
			$q_{E,m}^{(l+1)}[p]\leftarrow q_{E,m}^{\star}[p]$ and set $l=l+1$.
			\UNTIL{$\{q_{E,m}^{(l)}[p]\}$ converges}
			\STATE Recover $\{\mathbf w^{(r+1)}[p]\}_{p=1}^{P}$ from
			$\{\mathbf W^{\star}[p]\}_{p=1}^{P}$; apply Gaussian randomization
			if a rank-one solution is not obtained.
			\STATE Given $\{\mathbf w^{(r+1)}[p]\}_{p=1}^{P}$, construct the
			surrogate functions in (\ref{secrecy_surrogate}),
			(\ref{slot_surrogate}), and (\ref{spacing_surrogate}) at
			$\mathbf c^{(r)}$.
			\STATE Solve (P5) via CVX to obtain $\mathbf c^{(r+1)}$.
			\STATE Set $r=r+1$.
			\UNTIL{the average worst-case secrecy rate converges}
			\STATE \textbf{Output:} $\mathbf c^\star=\mathbf c^{(r)}$ and
			$\{\mathbf w^\star[p]\}_{p=1}^{P}
			=\{\mathbf w^{(r)}[p]\}_{p=1}^{P}$.
		\end{algorithmic}
	\end{algorithm}
	
	For the proposed AO algorithm, the beamforming update requires iteratively solving the convex problem (P3). For each time slot, (P3) involves an $N\times N$ positive semidefinite matrix variable and $M[p]$ LMIs of size $(N+1)\times(N+1)$. Hence, the computational complexity of the beamforming
	update can be approximately expressed as $\mathcal{O}\!\left(I_W\sum_{p=1}^{P}M[p]N^6\right)$, where $I_W$ denotes the number of inner iterations for updating the beamforming variables.
	
	The FA position subproblem (P5) can be formulated as a convex quadratically constrained quadratic program (QCQP) involving $2N$ real variables, subject to $P$ surrogate secrecy-rate constraints and $N(N-1)/2$ linearized antenna-spacing constraints. Its complexity can be approximately expressed as $\mathcal{O}\!\left(I_C(P+N(N-1)/2)(2N)^3\right)$, where $I_C$ denotes the number of iterations for updating ${\bf c}$. Therefore, the overall computational complexity is approximately
	\begin{align}
		{\cal O}({I_{{\rm{AO}}}}[{I_W}\sum\limits_{p = 1}^P M [p]{N^6} \!+\! {I_C}(P + \frac{{N(N \!-\! 1)}}{2}){(2N)^3}]),
	\end{align}
	where $I_{\rm AO}$ denotes the number of outer AO iterations.

	\section{Proposed DE-Based AO Algorithm}
	\label{pa4}
	
	In this section, we further employ a DE-based AO algorithm to solve (P1). Although the SCA-based AO algorithm has a clear mathematical structure and efficiently solves the approximated subproblems, its performance may depend on the initialization and the quality of the local approximations, especially
	for the highly non-convex FA position optimization with coupled inter-antenna spacing constraints. Motivated by this, we further propose a DE-based AO algorithm. As a derivative-free population-based search method, DE does not rely on gradient information or local convexification and provides stronger
	exploration capability over the non-convex feasible region. Therefore, the DE-based method serves as a complementary approach to the SCA-based method for FA position optimization.
	
	\subsection{DE Algorithm for Antenna Position Optimization}
	
	In the proposed DE-based AO algorithm, the beamforming subproblem is still solved by the SDR-based robust beamforming method developed in the previous subsection, while the FA position subproblem is directly handled by DE. Unlike SCA, DE is a derivative-free population-based method that evolves a population of candidate solutions over $G_{\max}$ generations \cite{zhang2025movable}. For a given set of beamforming vectors, the fitness function is defined based on the average worst-case secrecy rate as
	\begin{align}
		{\cal F}({\bf c})=
		\begin{cases}
			\bar C_s^{\rm wc}({\bf c}),
			& C_s^{\rm wc}({\bf c},t_p)\ge C_{\min},~\forall p\in{\cal P},\\
			-\infty, & \text{otherwise}.
		\end{cases}
		\label{DE_fitness}
	\end{align}
	The algorithm starts by randomly generating $N_{\rm pop}$ feasible individuals within the FA reconfiguration region while satisfying the minimum inter-antenna spacing constraint. The initial population is given by
	\begin{align}
		{\cal Q}^{(0)}
		=
		\left\{
		{\bf c}_1^{(0)},
		{\bf c}_2^{(0)},
		\ldots,
		{\bf c}_{N_{\rm pop}}^{(0)}
		\right\}.
		\label{p30}
	\end{align}
	The best individual is recorded as the initial elite solution
	\begin{align}
		{\bf c}_{\rm elite}^{(0)}
		=
		\arg\max_{{\bf c}_i^{(0)}\in{\cal Q}^{(0)}}
		{\cal F}({\bf c}_i^{(0)}).
	\end{align}
	For each generation $g=0,1,\ldots,G_{\max}-1$, mutation, crossover, feasibility repair, and greedy selection are sequentially performed as follows.
	
	\subsubsection{Mutation}
	For the $i$-th target individual, three mutually different indices $r_1$, $r_2$, and $r_3$, which are also different from $i$, are randomly selected. The mutant vector is generated as
	\begin{align}
		{\bf{v}}_i^{(g)} = {\bf{c}}_{{r_1}}^{(g)} + F({\bf{c}}_{{r_2}}^{(g)} - {\bf{c}}_{{r_3}}^{(g)}),
		\label{p31}
	\end{align}
	where $F$ denotes the mutation scale factor.
	
	\subsubsection{Crossover}
	Binomial crossover is performed between ${\bf v}_i^{(g)}$ and ${\bf c}_i^{(g)}$ to generate the trial vector ${\bf u}_i^{(g)}$. Its $k$-th component is given by
	\begin{align}
		{[{\bf{u}}_i^{(g)}]_k} = \left\{ {\begin{array}{*{20}{l}}
				{{{[{\bf{v}}_i^{(g)}]}_k},}&{{r_{i,k}} \le {C_R}\;{\rm{or}}\;k = {k_{{\rm{rand}}}},}\\
				{{{[{\bf{c}}_i^{(g)}]}_k},}&{{\rm{otherwise}},}
		\end{array}} \right.
		\label{p32}
	\end{align}
	where $r_{i,k}\sim{\cal U}(0,1)$, $C_R\in[0,1]$ is the crossover probability, and $k_{\rm rand}$ is a randomly selected index ensuring that at least one component is inherited from the mutant vector. After crossover, ${\bf u}_i^{(g)}$ may violate the FA reconfiguration region or minimum-spacing constraints. Therefore, a feasibility repair operation is performed before fitness evaluation. Each coordinate is first projected onto its feasible interval as
	\begin{align}
		{[{\bf{u}}_i^{(g)}]_k} \leftarrow \min \{ \max \{ {[{\bf{u}}_i^{(g)}]_k},{l_k}\} ,{u_k}\} ,
		\label{p33}
	\end{align}
	where $l_k$ and $u_k$ denote the lower and upper bounds of the $k$-th coordinate, respectively. For an antenna pair $(n,\tilde n)$ violating constraint (\ref{p4}), their positions are symmetrically displaced along their separating direction as
	\begin{align}
		{\bf c}_n
		\leftarrow
		{\bf c}_n
		+
		\frac{\Delta_{n\tilde n}}{2}
		{\bf d}_{n\tilde n},
		{\bf c}_{\tilde n}
		\leftarrow
		{\bf c}_{\tilde n}
		-
		\frac{\Delta_{n\tilde n}}{2}
		{\bf d}_{n\tilde n},
		\label{p34}
	\end{align}
	where
	\begin{align}
		{\bf d}_{n\tilde n}
		=
		\frac{
			{\bf c}_n-{\bf c}_{\tilde n}
		}{
			\|{\bf c}_n-{\bf c}_{\tilde n}\|_2
		},
		\Delta_{n\tilde n}
		=
		d_{\min}
		-
		\|{\bf c}_n-{\bf c}_{\tilde n}\|_2.
		\label{p35}
	\end{align}
	The projection and spacing-repair operations are repeatedly performed until the geometric constraints are satisfied.
	
	\subsubsection{Selection}
	After feasibility repair, the fitness of the trial individual is evaluated according to (\ref{DE_fitness}). Greedy selection is then performed as
	\begin{align}
		{\bf c}_i^{(g+1)}
		=
		\begin{cases}
			{\bf u}_i^{(g)},
			& {\cal F}({\bf u}_i^{(g)})
			\ge
			{\cal F}({\bf c}_i^{(g)}),\\
			{\bf c}_i^{(g)},
			& \text{otherwise}.
		\end{cases}
		\label{p36}
	\end{align}
	In addition, an elitism strategy is employed to preserve the best-so-far solution. Specifically,
	\begin{align}
		{\bf c}_{\rm elite}^{(g+1)}
		=
		\arg\max_{
			{\bf c}\in
			{\cal Q}^{(g+1)}
			\cup
			\{{\bf c}_{\rm elite}^{(g)}\}
		}
		{\cal F}({\bf c}).
	\end{align}
	The updated elite solution is retained in the next generation to prevent the loss of the best solution found so far.
	
	\subsection{The Convergence and Computational Complexity}
	
	\begin{algorithm}[!t]
		\caption{Robust DE-Based AO Algorithm for Solving (P1)}
		\label{alg:DE_P4}
		\begin{algorithmic}[1]
			\STATE \textbf{Initialize:} Set $r=0$, and choose feasible $\mathbf c^{(0)}$
			and $\{\mathbf w^{(0)}[p]\}_{p=1}^{P}$. Set $F$, $C_R$,
			$N_{\mathrm{pop}}$, $G_{\max}$, and $T_{\max}$.
			
			\REPEAT
			\STATE Given $\mathbf c^{(r)}$, solve (P3) to obtain
			$\{\mathbf W^{\star}[p]\}_{p=1}^{P}$.
			
			\STATE Recover $\{\mathbf w^{(r+1)}[p]\}_{p=1}^{P}$ from
			$\{\mathbf W^{\star}[p]\}_{p=1}^{P}$; apply Gaussian randomization
			if a rank-one solution is not obtained.
			
			\STATE Set $g=0$ and initialize the population
			$\mathcal Q^{(0)}=\{\mathbf c_i^{(0)}\}_{i=1}^{N_{\mathrm{pop}}}$,
			with $\mathbf c^{(r)}$ included as one individual.
			
			\STATE Repair all individuals using (\ref{p33})--(\ref{p35}) and
			evaluate their fitness values $\mathcal F(\mathbf c_i^{(0)})$.
			
			\STATE Set
			$\mathbf c_{\mathrm{elite}}^{(0)}
			=\arg\max_{\mathbf c_i^{(0)}\in\mathcal Q^{(0)}}
			\mathcal F(\mathbf c_i^{(0)})$.
			
			\REPEAT
			\FOR{$i=1$ to $N_{\mathrm{pop}}$}
			\STATE Randomly select mutually different indices
			$r_1$, $r_2$, and $r_3$, with
			$r_1,r_2,r_3\neq i$.
			
			\STATE Mutation:
			$\mathbf v_i^{(g)}
			=\mathbf c_{r_1}^{(g)}
			+F\bigl(\mathbf c_{r_2}^{(g)}
			-\mathbf c_{r_3}^{(g)}\bigr)$.
			
			\STATE Crossover: Generate $\mathbf u_i^{(g)}$ from
			$\mathbf v_i^{(g)}$ and $\mathbf c_i^{(g)}$
			according to (\ref{p32}).
			
			\STATE Repair $\mathbf u_i^{(g)}$ using
			(\ref{p33})--(\ref{p35}) until the geometric constraints
			are satisfied.
			
			\STATE Evaluate $\mathcal F(\mathbf u_i^{(g)})$.
			
			\IF{$\mathcal F(\mathbf u_i^{(g)})
				\ge \mathcal F(\mathbf c_i^{(g)})$}
			\STATE $\mathbf c_i^{(g+1)}=\mathbf u_i^{(g)}$.
			\ELSE
			\STATE $\mathbf c_i^{(g+1)}=\mathbf c_i^{(g)}$.
			\ENDIF
			\ENDFOR
			
			\STATE Update
			$\mathbf c_{\mathrm{elite}}^{(g+1)}
			=\arg\max_{\mathbf c\in
				\mathcal Q^{(g+1)}\cup\{\mathbf c_{\mathrm{elite}}^{(g)}\}}
			\mathcal F(\mathbf c)$.
			
			\STATE Retain $\mathbf c_{\mathrm{elite}}^{(g+1)}$
			in the next generation.
			
			\STATE Set $g=g+1$.
			\UNTIL{$g=G_{\max}$}
			
			\STATE Set
			$\mathbf c^{(r+1)}=\mathbf c_{\mathrm{elite}}^{(G_{\max})}$.
			
			\STATE Set $r=r+1$.
			\UNTIL{$r=T_{\max}$}
			
			\STATE \textbf{Output:}
			$\mathbf c^\star=\mathbf c^{(r)}$ and
			$\{\mathbf w^\star[p]\}_{p=1}^{P}
			=\{\mathbf w^{(r)}[p]\}_{p=1}^{P}$.
		\end{algorithmic}
	\end{algorithm}
	
	For the DE-based AO algorithm, the beamforming subproblem is solved using the SDR-based robust beamforming method described in the preceding subsection, while the FA position subproblem is solved via DE. Elitism preserves the best feasible individual during the DE search, and each outer AO update is accepted only if it preserves feasibility and does not decrease the average worst-case secrecy rate. Starting from a feasible initialization, the resulting objective sequence is therefore nondecreasing. Since this sequence is bounded above under the finite transmit-power constraint, it converges. The detailed procedure is summarized in \textbf{Algorithm~2}.
	
	For the DE-based AO algorithm, the beamforming update has the same computational complexity as that of the SCA-based AO algorithm, i.e.,
	$\mathcal{O}\!\left(I_W\sum_{p=1}^{P}M[p]N^6\right)$. For the FA position update, DE performs $G_{\max}$ generations with $N_{\mathrm{pop}}$ individuals. The dominant computational cost arises from evaluating the fitness values of all candidate solutions over $P$ time slots and the corresponding eavesdropping links, resulting in a complexity of approximately $\mathcal{O}\!\left(G_{\max}N_{\mathrm{pop}}N\left(P+\sum_{p=1}^{P}M[p]\right)\right)$. Therefore, the overall computational complexity can be approximated as ${\cal O}\big({T_{\max }}[{I_W}\sum\nolimits_{p = 1}^P M [p]{N^6} + {G_{\max }}{N_{{\rm{pop}}}}N(P + \sum\nolimits_{p = 1}^P M [p])])$.
	
	\section{Numerical Results}
	\label{pa5}
	
	This section provides a comprehensive simulation-based evaluation of the proposed system and the developed algorithms. Unless otherwise specified, the system operates at $12\ {\rm GHz}$. The LEO satellites are deployed at an altitude of $550\ {\rm km}$, with the Earth's radius set to $6371\ {\rm km}$ and the orbital inclination angle set to $50^\circ$. The latitude of the GS is $\Theta_G=40^\circ$. The radius of the circular aperture of the satellite antenna is set to $r=3\lambda$, and the aperture efficiency is $\eta=0.5$, yielding a maximum antenna gain of $D_0=\eta(2\pi r/\lambda)^2\approx22.5\ {\rm dBi}$. The atmospheric and rain attenuation are set to $A_{jk}^{\rm atm}(t)=0.5\ {\rm dB}$ and $A_{jk}^{\rm rain}(t)=2\ {\rm dB}$, respectively. Without loss of generality, the initial propagation phase is set to $\phi_{0,jk}=0$. The minimum inter-element spacing of the FA array is $d_{\min}=0.5\lambda$, and the FA moving region is ${\cal C}=3\lambda\times3\lambda$. The GS transmit power is $40\ {\rm dBm}$, and the noise power is $-148\ {\rm dBm}$. The minimum required secrecy rate is $C_{\min}=0.01\ {\rm bit/s/Hz}$, and the path-loss exponent is $\alpha=2$. The number of time slots is $P=30$. The smoothing parameter is set to $\mu=0.05$. For the DE-based AO algorithm, the maximum number of outer AO iterations is $T_{\max}=100$, the population size is $N_{\rm pop}=50$, the maximum number of generations is $G_{\rm max}=1000$, the mutation scale factor is $F=0.9$, and the crossover probability is $C_R=0.9$. To characterize the ECSI uncertainty level, we define the channel uncertainty ratio as $
	\delta_E=\frac{\varepsilon_{E,m}(t_p)}{\|\hat{\mathbf h}_{E,m}(\mathbf c,t_p)\|_2}$. Unless otherwise specified, we set \(\delta_E=0.01\).
		\begin{figure}[!t]
		\centering
		\includegraphics[width=0.4\textwidth]{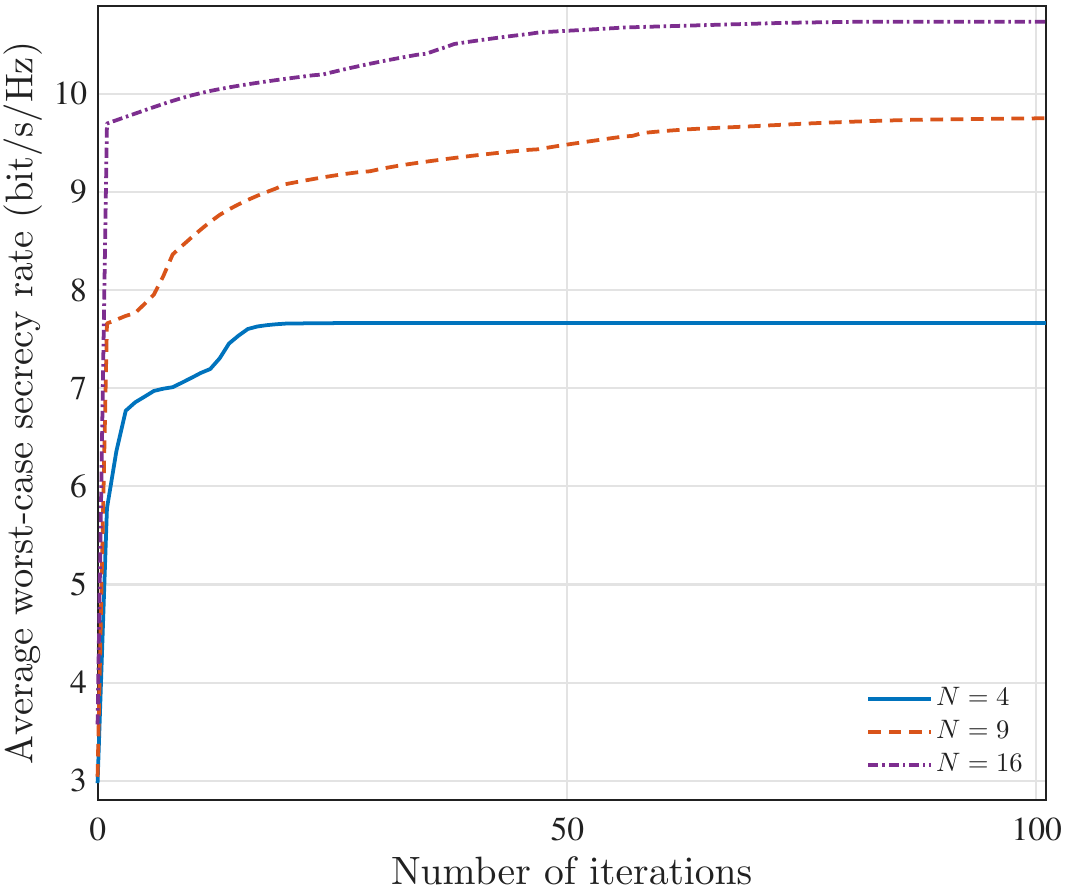}
		\caption{Convergence behavior of the proposed SCA-based AO algorithm.}
		\label{fig3}
	\end{figure}
	\begin{figure}[!t]
		\centering
		\includegraphics[width=0.4\textwidth]{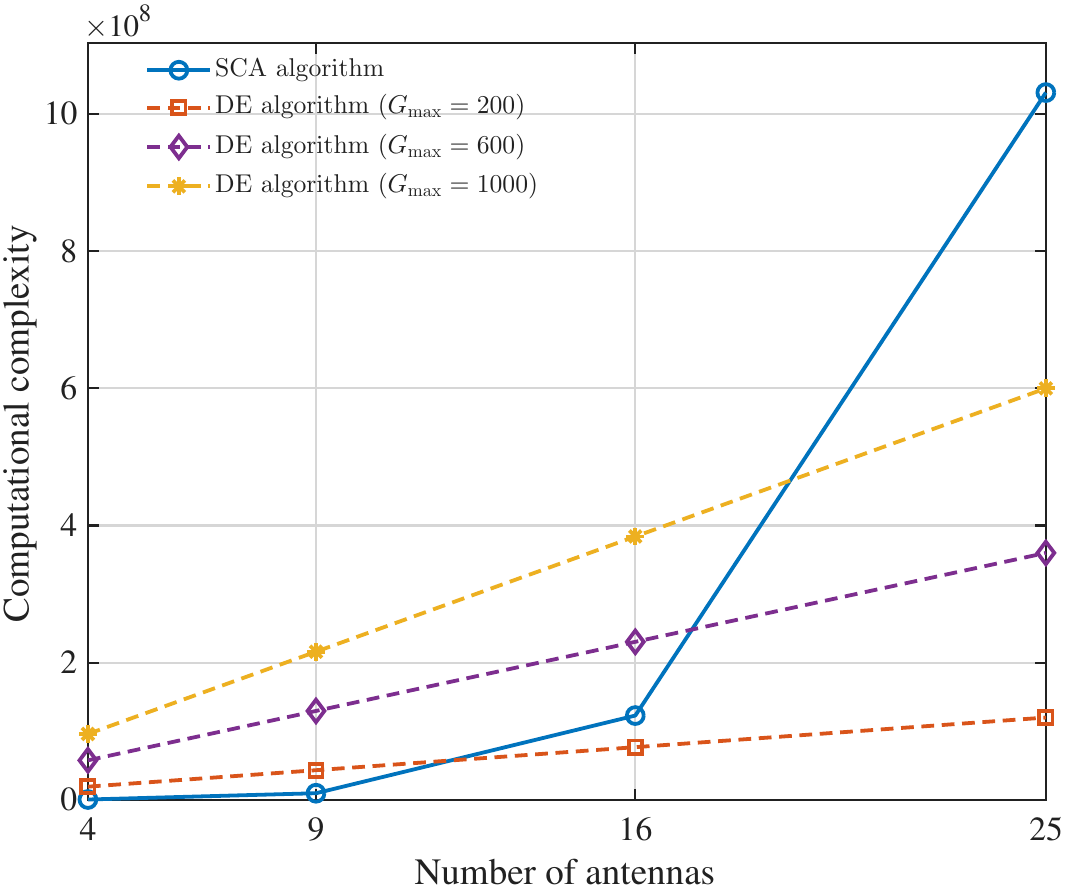}
		\caption{Complexity comparison of different algorithms.}
		\label{fig5}
	\end{figure}
	\begin{figure}
		\centering
		\includegraphics[width=0.4\textwidth]{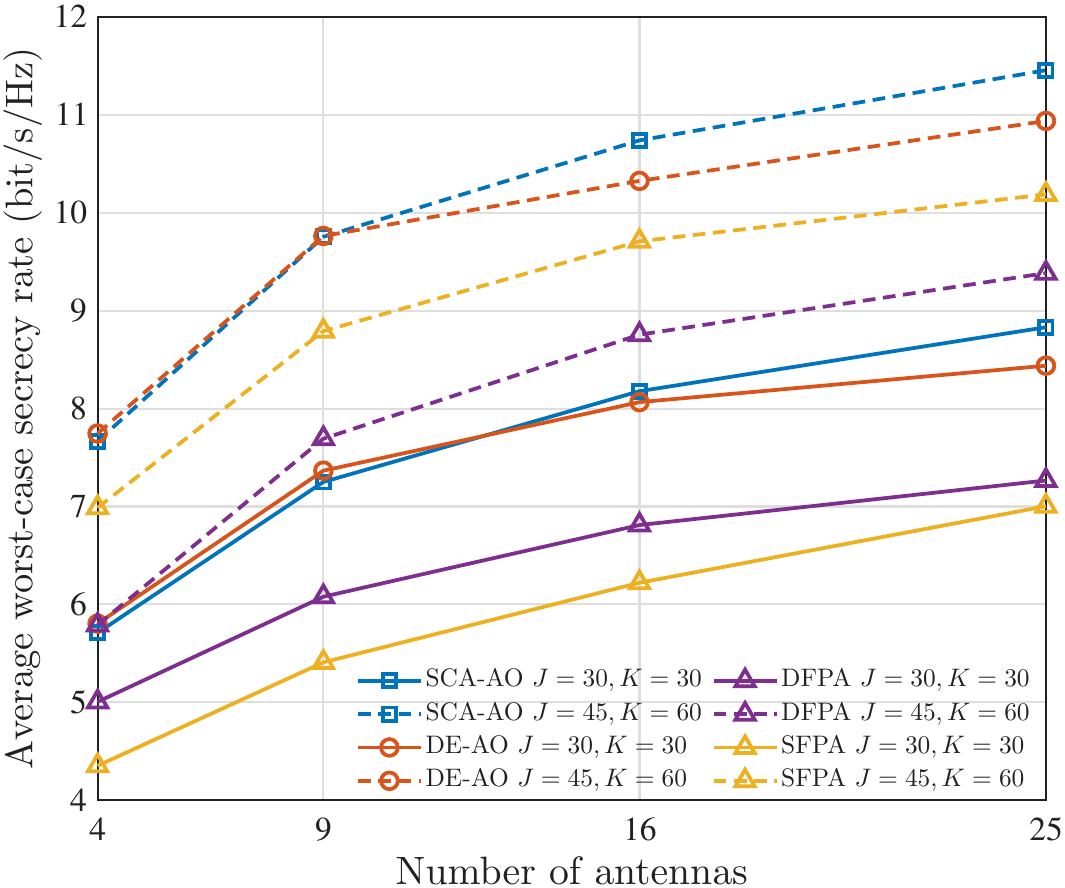}
		\caption{Average worst-case secrecy rate under different antenna schemes.}
		\label{fig6}
	\end{figure}

	To facilitate a clear and fair comparison, the following benchmark schemes are adopted:
	1) {\bf SCA-AO}, where the FA positions are optimized using SCA;
	2) {\bf DE-AO}, where the FA positions are optimized using DE;
	3) {\bf DFPA}, where the antenna positions are fixed with an inter-element spacing of $\lambda/2$, and only the transmit beamforming vectors are optimized;
	4) {\bf SFPA}, where the antenna positions are fixed with an inter-element spacing of $\lambda$, and only the transmit beamforming vectors are optimized.
	For both FA-based schemes, the antenna positions are initialized on a uniform planar grid with an adjacent-element spacing of $\lambda/2$.
	
	Fig. \ref{fig3} illustrates the convergence behavior of the SCA-based AO algorithm for different numbers of FAs under a fixed constellation configuration of $J=45$ and $K=60$. It can be observed that all curves increase monotonically and gradually converge, demonstrating the effectiveness and convergence behavior of the proposed algorithm. Moreover, the average worst-case secrecy rate increases significantly with the number of FAs. This is because a larger antenna array provides higher beamforming gain toward the serving satellite and more spatial flexibility to suppress signal leakage toward potential eavesdroppers, thereby improving secrecy performance.

	Fig. \ref{fig5} compares the computational complexity of the SCA- and DE-based FA position optimization algorithms versus the number of antennas. The complexity of SCA increases rapidly with the antenna number due to the growing dimension and number of constraints in the convex optimization subproblems. In contrast, the complexity of DE grows more slowly with the antenna number and increases linearly with the maximum number of generations $G_{\max}$. For a small number of antennas, SCA exhibits lower computational complexity than DE. However, as the antenna number increases, DE becomes more computationally attractive. In particular, when $N=25$, DE achieves lower theoretical complexity than SCA even with $G_{\max}=1000$. These results indicate that the DE-based method offers a computational advantage for FA position optimization in large antenna arrays.

	\begin{figure*}[!t]
		\centering
		
		\begin{minipage}{0.30\textwidth}
			\centering
			\includegraphics[width=\linewidth]{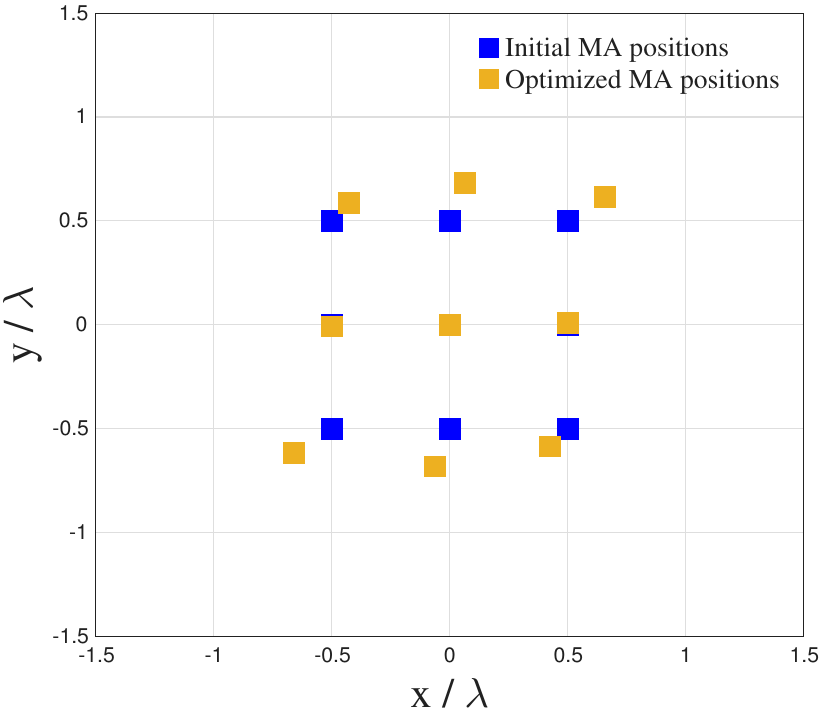}
			\\[2mm]
			(a) $J=15$, $K=20$
		\end{minipage}
		\hfill
		\begin{minipage}{0.30\textwidth}
			\centering
			\includegraphics[width=\linewidth]{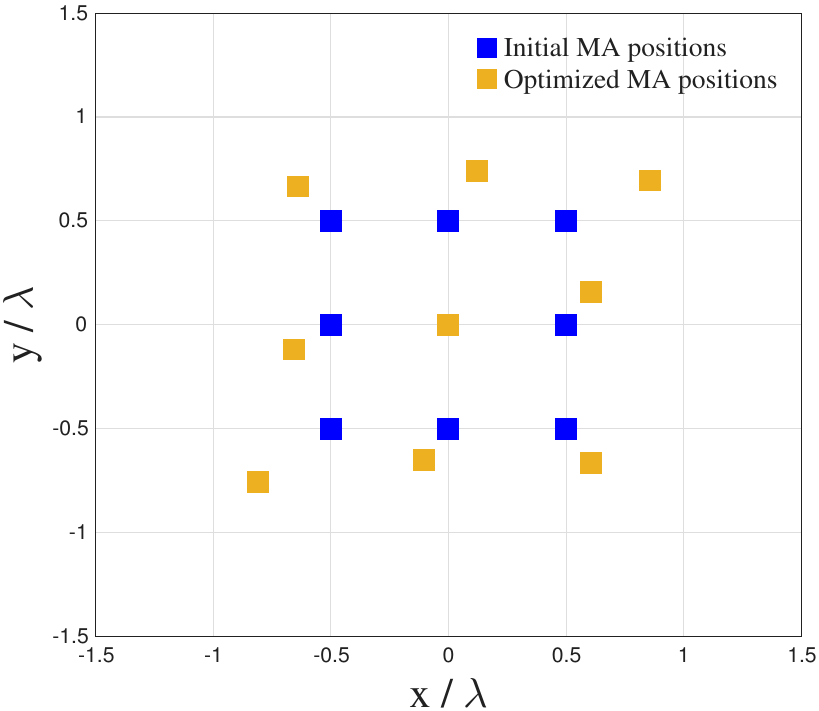}
			\\[2mm]
			(b) $J=30$, $K=30$
		\end{minipage}
		\hfill
		\begin{minipage}{0.30\textwidth}
			\centering
			\includegraphics[width=\linewidth]{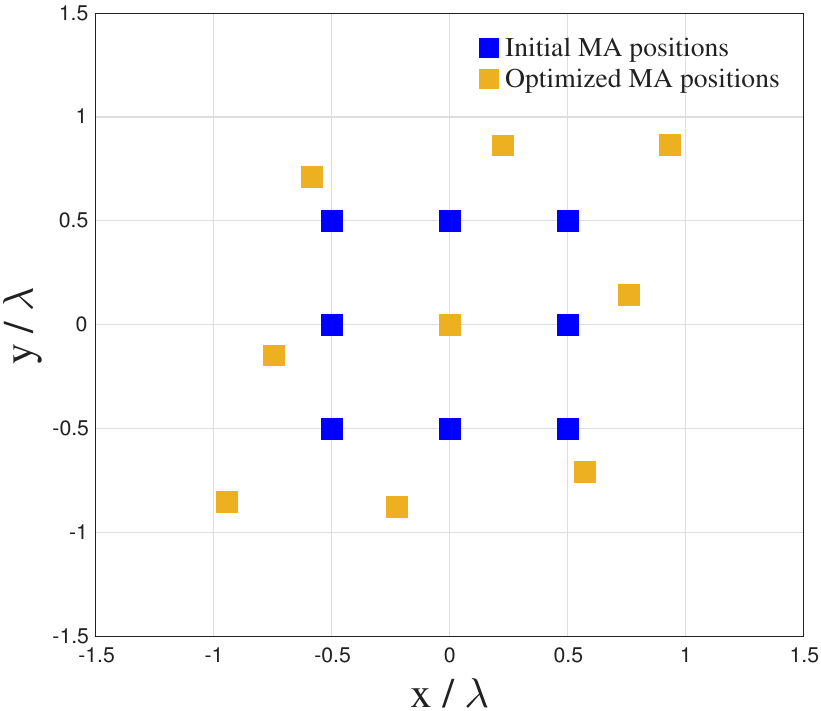}
			\\[2mm]
			(c) $J=45$, $K=60$
		\end{minipage}
		\caption{FA positions before and after optimization under different LEO satellite constellation configurations.}
		\label{fig7}
	\end{figure*}
	
	\begin{figure*}[]
		\centering
		
		\begin{minipage}{0.30\textwidth}
			\centering
			\includegraphics[width=\linewidth]{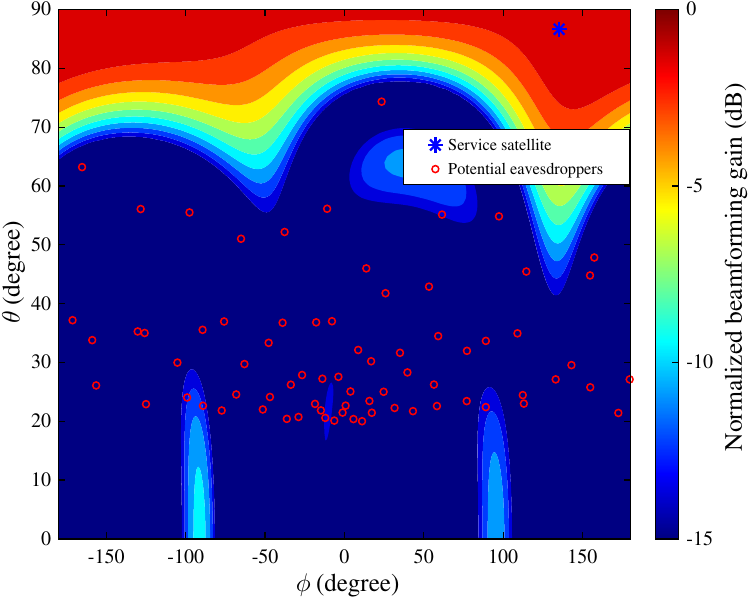}
			\\[2mm]
			(a) FA, SCA-AO, $p=2$
		\end{minipage}
		\hfill
		\begin{minipage}{0.30\textwidth}
			\centering
			\includegraphics[width=\linewidth]{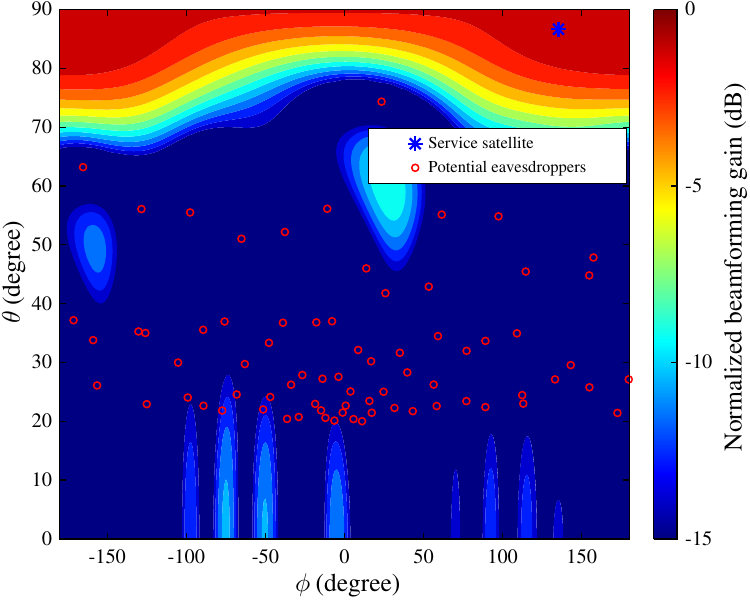}
			\\[2mm]
			(b) FA, DE-AO, $p=2$
		\end{minipage}
		\hfill
		\begin{minipage}{0.30\textwidth}
			\centering
			\includegraphics[width=\linewidth]{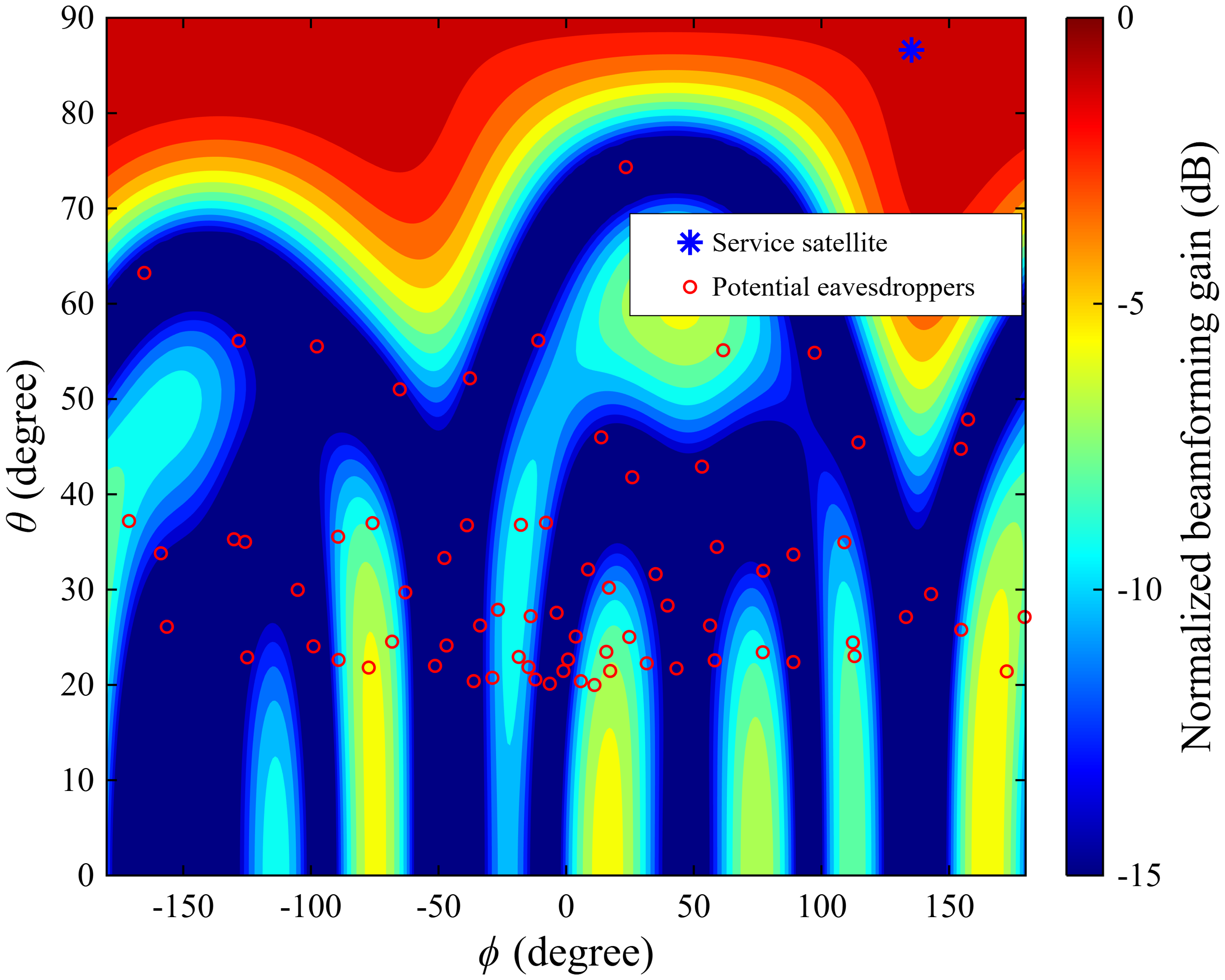}
			\\[2mm]
			(c) DFPA, $p=2$
		\end{minipage}
		
		\vspace{2mm}
		
		\begin{minipage}{0.30\textwidth}
			\centering
			\includegraphics[width=\linewidth]{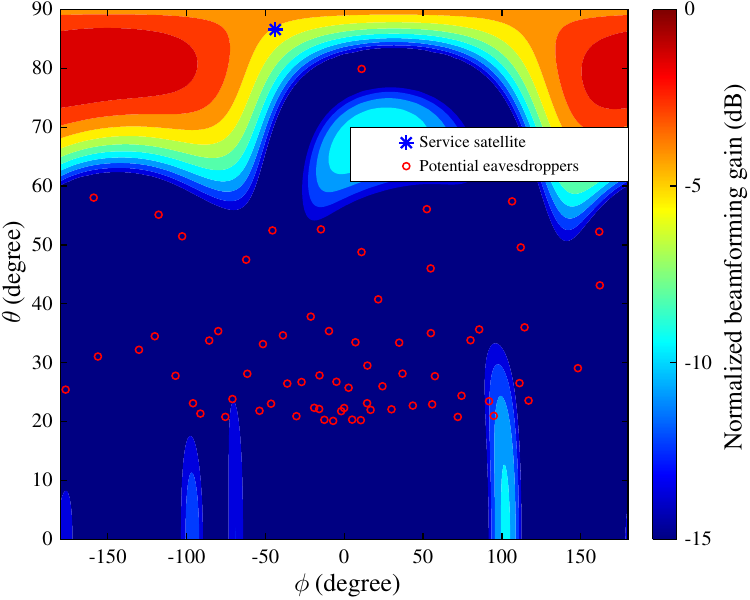}
			\\[2mm]
			(d) FA, SCA-AO, $p=29$
		\end{minipage}
		\hfill
		\begin{minipage}{0.30\textwidth}
			\centering
			\includegraphics[width=\linewidth]{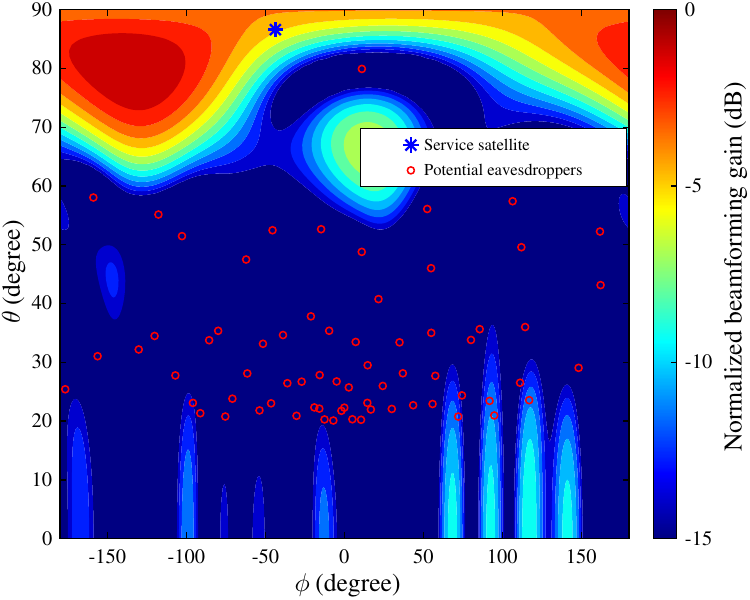}
			\\[2mm]
			(e) FA, DE-AO, $p=29$
		\end{minipage}
		\hfill
		\begin{minipage}{0.30\textwidth}
			\centering
			\includegraphics[width=\linewidth]{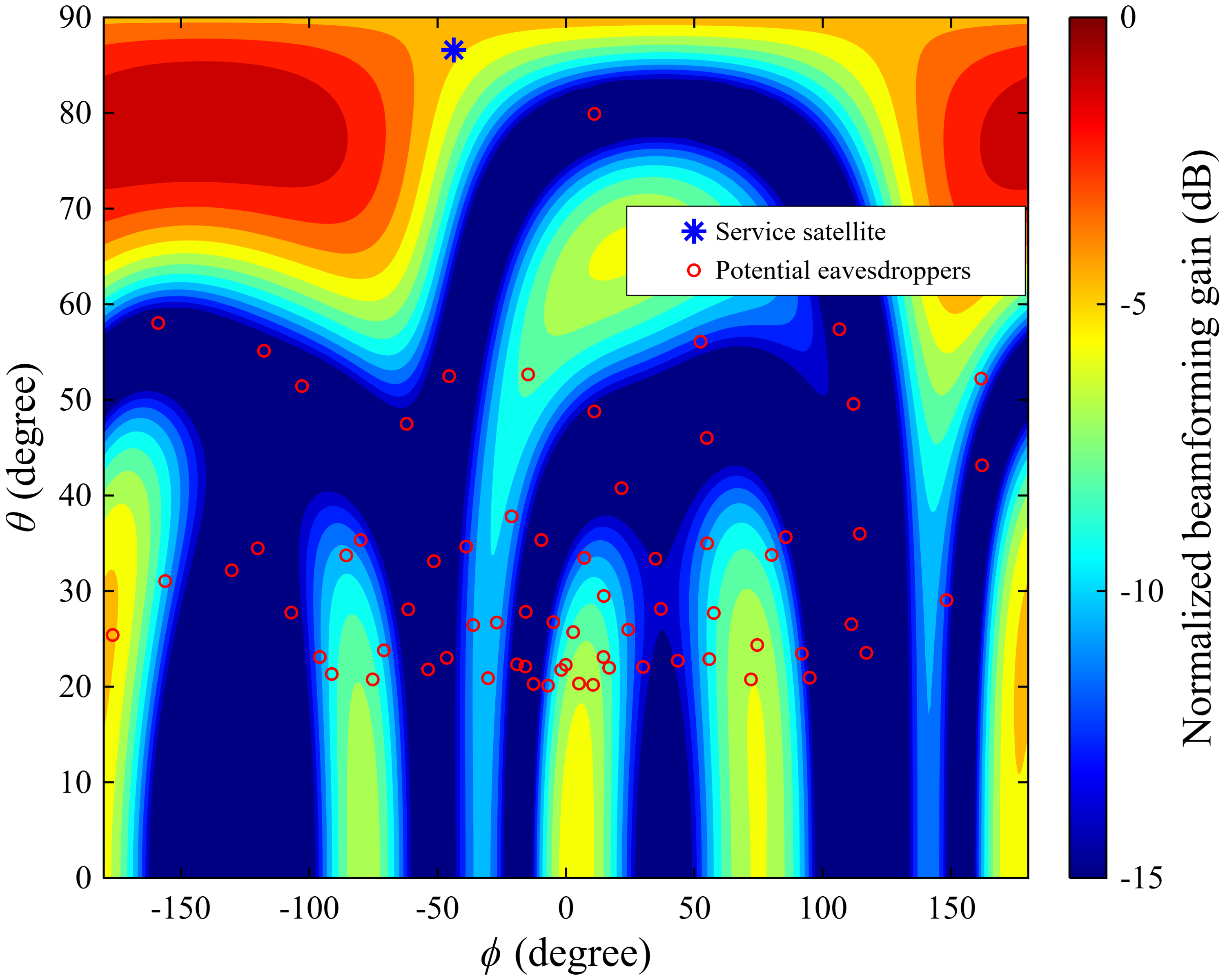}
			\\[2mm]
			(f) DFPA, $p=29$
		\end{minipage}
		\caption{Normalized beamforming gain patterns achieved by different schemes at different time slots.}
		\label{fig8}
	\end{figure*}
	
	Fig. \ref{fig6} illustrates the average worst-case secrecy rate versus the number of antennas under two LEO satellite constellation configurations, namely $J=30$, $K=30$, and $J=45$, $K=60$. As observed, the secrecy rate increases with the number of antennas for all schemes, since a larger antenna array provides higher beamforming gain and more spatial flexibility to enhance the legitimate link while suppressing information leakage. Moreover, DE-AO outperforms SCA-AO when the antenna number is small, benefiting from its global search capability in a relatively low-dimensional solution space. However, SCA-AO achieves better performance as the antenna number increases, since it can efficiently exploit the local problem structure, whereas the search efficiency of DE decreases in higher-dimensional optimization problems. For the fixed-position antenna schemes, DFPA outperforms SFPA when $J=30$ and $K=30$, whereas the opposite trend is observed when $J=45$ and $K=60$. This difference can be attributed to the spatial distributions of the serving satellite and potential eavesdroppers. In a larger constellation, the increased satellite density may result in smaller angular separations between the serving satellite and nearby eavesdroppers, requiring higher angular resolution for effective spatial discrimination. In this case, SFPA benefits from its larger array aperture, which improves angular resolution and enables better suppression of information leakage. In contrast, for the smaller constellation, DFPA achieves better performance by avoiding the increased grating-lobe effects associated with larger inter-element spacing. Moreover, both FA-based schemes achieve higher secrecy rates than DFPA and SFPA for the same constellation configuration, highlighting the benefits of joint antenna position and beamforming optimization. It is also observed that the constellation with $J=45$ and $K=60$ achieves higher secrecy rates in the considered scenarios. Although a larger constellation increases the number of potential eavesdroppers, it also provides more opportunities to establish a stronger legitimate link. Therefore, the secrecy performance depends on the relative channel conditions and angular distributions of the serving satellite and potential eavesdroppers, rather than the constellation size alone.
		\begin{figure}[!t]
		\centering
		\includegraphics[width=0.4\textwidth]{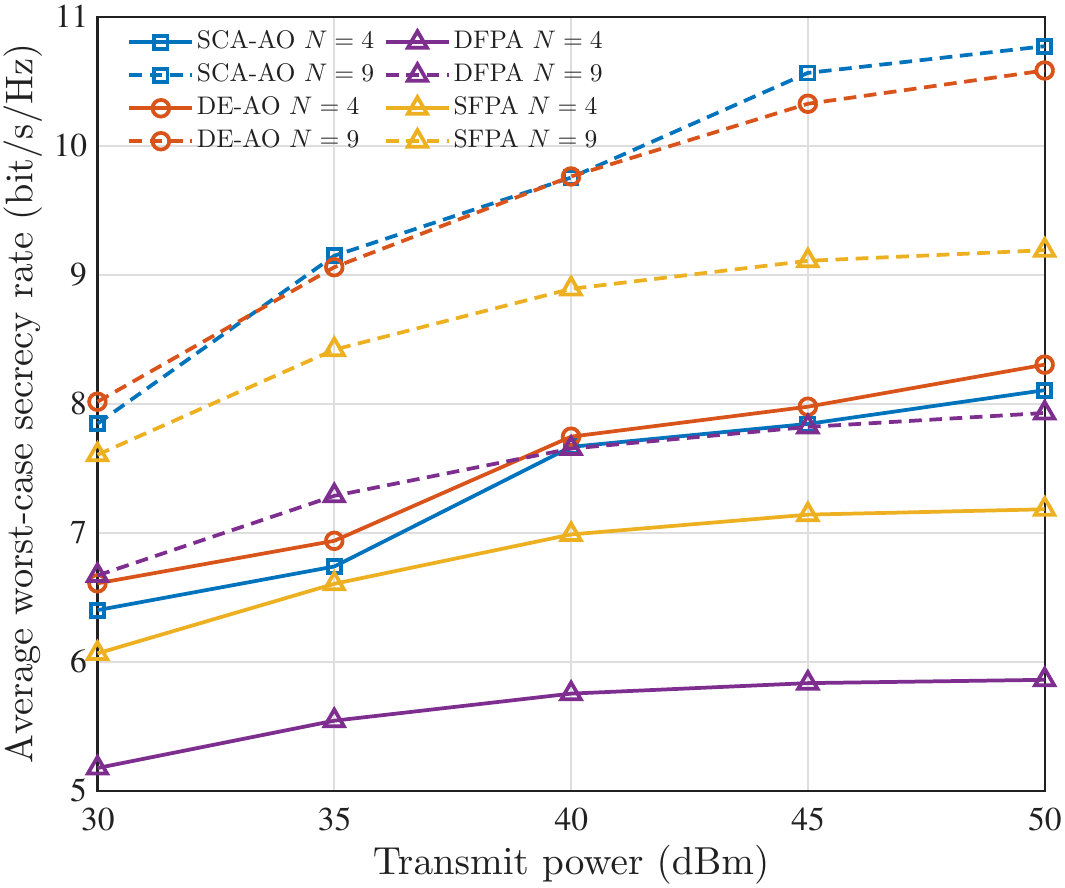}
		\caption{Average worst-case secrecy rate versus the transmit power under different schemes.}
		\label{fig9}
	\end{figure}
		\begin{figure}[!t]
		\centering
		\includegraphics[width=0.4\textwidth]{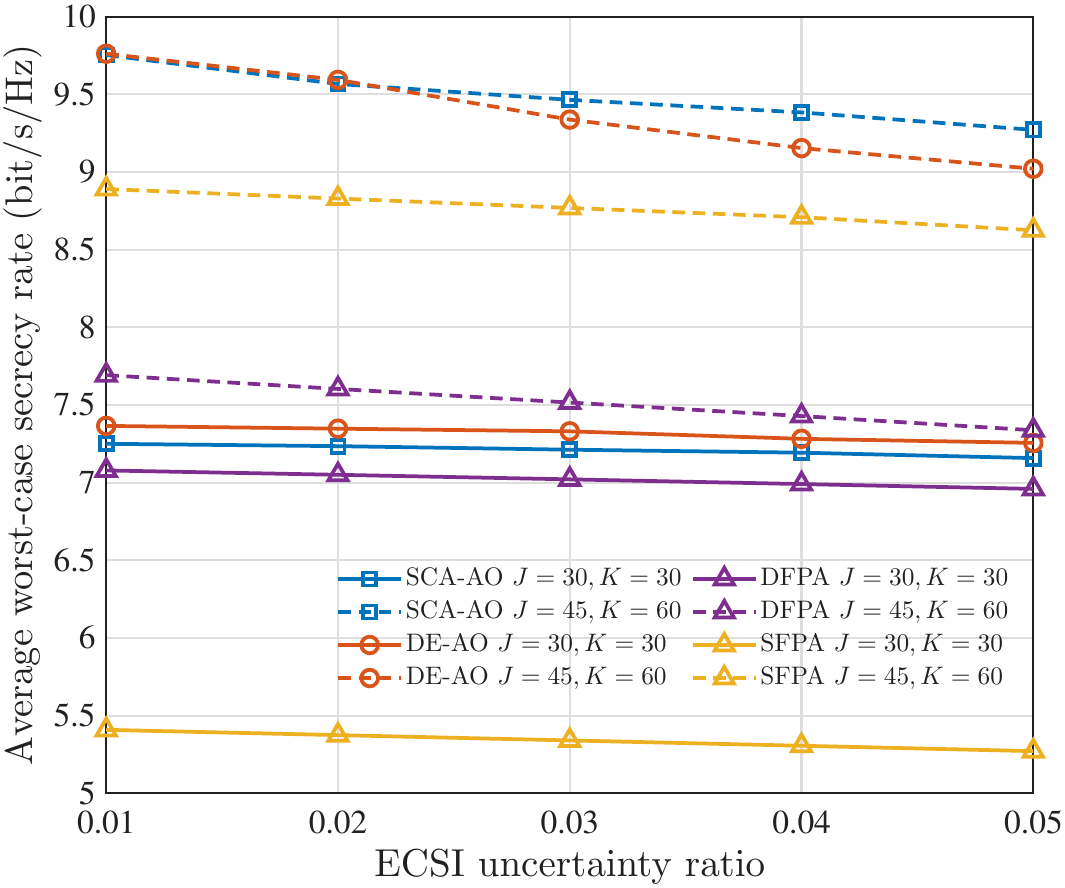}
		\caption{Average worst-case secrecy rate versus $\delta_E$ under different schemes.}
		\label{fig10}
	\end{figure}

	Fig. \ref{fig7} illustrates the optimized FA positions under different satellite constellation configurations, with $N=9$ and $P_{\max}=40$ dBm. As the constellation size increases, the FA elements tend to spread toward the boundary of the feasible region, resulting in a larger effective array aperture. This behavior can be attributed to the increased number and changing angular distribution of potential eavesdropping satellites, which impose greater demands on spatial discrimination. A larger array aperture provides higher angular resolution and greater flexibility in beam pattern design, allowing the GS to better distinguish the serving satellite from potential eavesdroppers and suppress information leakage.

	Fig.~\ref{fig8} compares the normalized beamforming gain patterns achieved by different schemes at time slots $p=2$ and $p=29$, with $J=45$, $K=60$, and $P_{\max}=40$ dBm. Due to orbital motion, the angular positions of the serving satellite and potential eavesdroppers change over time, resulting in different beam patterns across the two slots. It can be observed that both FA-based schemes maintain high beamforming gains toward the serving satellite while effectively suppressing signal radiation in most eavesdropping directions. In particular, SCA-AO produces relatively smooth low-gain regions, whereas DE-AO exhibits several localized sidelobes, especially at low elevation angles. In contrast, the DFPA scheme generates more pronounced sidelobes across a wider angular range, some of which overlap with potential eavesdropping directions. This is because the FA schemes can jointly optimize antenna positions and beamforming weights, providing additional spatial flexibility to improve directional selectivity and suppress information leakage. These results demonstrate the advantages of FA-aided beamforming in adapting to the time-varying spatial distributions of LEO satellites.

	Fig.~\ref{fig9} illustrates the average worst-case secrecy rate versus the maximum transmit power under different antenna configurations when $J=45$ and $K=60$. The secrecy rate improves with increasing transmit power across all schemes, but the performance gains become less pronounced at higher power levels. This is because increasing the transmit power enhances legitimate reception but may also increase information leakage toward potential eavesdroppers. Both FA-based schemes consistently outperform DFPA and SFPA. Among the fixed-position schemes, SFPA achieves higher secrecy rates than DFPA for both antenna configurations. This can be attributed to the larger array aperture of SFPA, which provides higher angular resolution and improves spatial discrimination between the serving satellite and nearby eavesdroppers. Regarding the proposed algorithms, DE-AO outperforms SCA-AO when $N=4$, while the two algorithms achieve comparable secrecy performance when $N=9$. Furthermore, increasing the antenna number from $N=4$ to $N=9$ significantly improves the secrecy performance of all schemes by providing additional spatial flexibility for enhancing the legitimate link and suppressing information leakage.

	Fig.~\ref{fig10} illustrates the average worst-case secrecy rate versus the ECSI uncertainty ratio under different satellite constellation configurations when $N=9$. As observed, the secrecy rate decreases with increasing ECSI uncertainty for all schemes. This is because a larger uncertainty ratio expands the norm-bounded ECSI uncertainty region, allowing more adverse eavesdropping channel realizations to be considered in the worst-case evaluation, thereby increasing the potential information leakage. Nevertheless, both FA-based schemes consistently outperform DFPA and SFPA, demonstrating the benefits of joint antenna position and robust beamforming optimization under imperfect ECSI. Moreover, the constellation with $J=45$ and $K=60$ achieves higher secrecy rates than that with $J=30$ and $K=30$ across all considered schemes and uncertainty levels. A larger constellation increases both the number of potential eavesdroppers and the opportunities to establish a stronger legitimate link. Thus, secrecy performance depends on their relative channel conditions and angular distributions rather than constellation size alone.
	
	\section{Conclusion}
	\label{pa6}

	In this paper, we investigated robust secure uplink transmission in time-varying LEO satellite constellations by deploying a planar FA array at the GS. Considering imperfect ECSI with norm-bounded errors, we formulated an average worst-case secrecy rate maximization problem by jointly optimizing the FA positions and transmit beamforming. To tackle the resulting non-convex problem, an AO framework was developed, where the beamforming subproblem was addressed using SDR, while the FA position optimization subproblem was solved using SCA-based and DE-based methods. Numerical results demonstrated that the proposed FA-aided schemes consistently outperform the conventional FPA scheme by exploiting additional spatial flexibility to enhance legitimate reception and suppress information leakage. Moreover, DE-AO achieved slightly better performance with a small number of antennas, whereas SCA-AO exhibited superior performance as the problem dimension increased. These results highlight the potential of FA-aided beamforming for enhancing secure uplink transmission in dynamic LEO satellite networks.

	\bibliographystyle{IEEEtran}
	\bibliography{ref1}
\end{document}